\documentclass[useAMS, usenatbib, fleqn]{mn2e}
\usepackage{graphicx}

\def\apss{ Astrophys. Space. Sci.}
\def\apj{APJ}
\def\apjl{APJ, Part 2 - Letters}
\def\aap{ Ann. Phys. (Paris)}
\def\araa{Annual Review of Astronomy \& Astrophysics}
\def\mnras{MNRAS}
\def\nat{Nature}
\def\pasj{Publ. of the Astronomical Society of Japan}
\def\prd{Physical Review D}
\def\physrep{Phys. Rep.}

\title{Quasi-Periodic Flares from Star-Accretion Disc Collisions}

\author[Lixin (Jane) Dai, Steven V. Fuerst, \& Roger Blandford]{Lixin Dai$^{1}$\thanks{E-mail:
cosimo@stanford.edu (JD)}, Steven V. Fuerst$^{2}$ \thanks{E-mail: sfuerst@stanford.edu (SF)} and Roger Blandford$^{3}$\thanks{E-mail: rdb3@stanford.edu (RB)} \\
$^{1,2,3}$Kavli Institute for Particle Astrophysics and Cosmology, Stanford University, Menlo Park, CA 94025, USA\\}

\begin{document}

\pagerange{\pageref{firstpage}--\pageref{lastpage}} \pubyear{2009}

\label{firstpage}

\maketitle

\begin{abstract}

We present simulated results of quasi-periodic flares generated by the inelastic collisions of a star bound to a super-massive black hole (SMBH) and its attendant accretion disc. We show that the behavior of the quasi-periodicity is affected by the mass and spin of the black hole and the orbital elements of the stellar orbit. We also evaluate the possibility of extracting useful information on these parameters and verifying the character of the Kerr metric from such quasi-periodic signals. Comparisons are made with the observed optical outbursts of OJ287, infrared flares from the Galactic center and X-ray variability in RE J1034+396.

\end{abstract}


\section{Introduction}

One of the important developments in our understanding of galactic nuclei has been the realization that stars are intimately involved in the dynamics of accretion discs. Direct evidence for this comes from observations of Sgr $A^\star$, where three-dimensional stellar orbits are determined within $10^{15}$cm of the compact radio source with no preferred orientation \citep {Ghez:05, Genzel:07, Gillessen}. These observations lend support to the notion that stars can actually be formed from accreting gas in a disc. This realization leads to consideration of the dynamical interaction between stars and discsand the observable consequences of this interaction.

\begin{figure}
 \centering
 \includegraphics[width=3in]{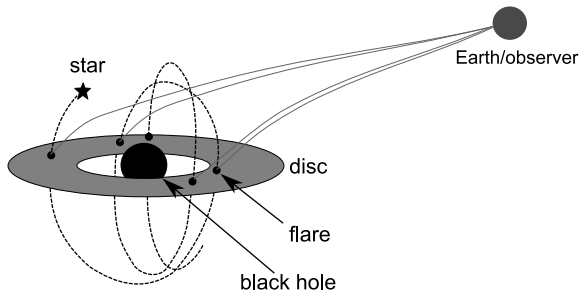}
 \caption{A star bound in the gravitational potential of the black hole: it orbits around the black hole, hits the accretion disc, and produces flares repeatedly.} 
 \label{model}
\end{figure}

The passage of a star through an accretion disc has been the focus of many studies. Stars can influence the disc by imposing a drag and removing its angular momentum \citep[e.g.][]{Ostriker:83, Norman:83}, heating \citep[e.g.][]{Perry:93}, and depositing mass \citep[e.g.][]{Armitage:96}. On the other hand, the distribution of the stellar orbits can be altered by the disc \citep[e.g.][]{Syer:91, Rauch:95, Karas:01}. \citet{Zentsova:83} showed that each passage through a disc results in the appearance of a bright hot spot on the disc's surface. \citet{Perry:93} related this passage to a UV bump superimposed upon the continuum spectrum emitted by an optically thick disc. \citet{Zurek} showed that star-disc collisions can leave ``star tails" above the disc and produce broad line emission. \citet{Nayakshin:04} used the passage to explain a large magnitude and short duration X-ray flare at the Galactic Centre \citep{Baganoff:01}. However, a bound star will pass many times through a disc, each passage leading to a flare, as shown in Fig.~\ref{model}. The timing of these flares can therefore be a measure of the elements of the stellar orbit and, when the orbit is relativistic, it can provide a probe of strong-field gravity. In this paper, we consider what information could, in principle, be derived from careful observations of a single star in an individual relativistic orbit about a black hole and its attendant disc.

In classical mechanics, when a star orbits a massive black hole, the stellar trajectory is a bound ellipse with the position of the heavier object at the focus. However, in the strong gravitational potential well of black hole, a star's trajectory will be more complicated. The three-dimensional orientation of each stellar orbit differs due to two general relativity effects. First, the peribothon, or closest approach to the hole, precesses on the orbital plane. Second, if the black hole is spinning, due to the dragging of inertial frames, the orbital plane itself will precess about the spin of the hole; henceforth, the orbit becomes even more complex \citep[e.g.][]{Misner:73}.

Not only massive particles but also massless photons' trajectories are altered by space-time curvature. Relativistic ray-tracing has been tackled both analytically and numerically. \citet{Cunningham73} use this method to derive the transfer function that allows one to image objects near black holes. The propagation of photons from the accretion disc around black holes has been well studied using semi-analytical methods and direct numerical methods by many people (see for example, \citet{Piran:83, Laor:91, Kojima:91, Rauch:92, Agol:97, Dab:97, Bromley:97, Pariev:98, Rey:99, Fuerst:04}, and references therein).

Combining the stellar trajectory, the star-disc collision model, and ray-tracing the photons from the flares, the time behavior and flux of the flares caused by star-disc collision can be modelled.  These signals exhibit quasi-periodicity for two reasons: (1) due to the precession of a stellar orbit, intrinsic time separations between two consecutive star-disc collisions are different, and (2) because the star hits the accretion disc at different ($r, {\phi}$) coordinates on the equatorial plane each time, the light-crossing time for the signals is also different.

This paper is organized as follows. In section 2, we give the assumptions and methodology, describe the stellar trajectories, construct our collision model, and discuss the method of ray-tracing. In section 3, we present simulated results for the quasi-periodic signals, showing in detail how the quasi-periodicity depends on the parameters of the black hole and the star's orbit. In section 4,  we discuss the possibility and method of extracting useful information from the signals. In section 5, we relate our model to observations of quasi-periodic behaviors in AGN. Further implications are briefly discusses in the last section.

\label{lastpage}

\section{Model and Methodology}

\subsection{Assumptions and methodology}

The prime assumption in this paper is that the gas particles in the accretion disc and the star orbiting the SMBH can be treated as massive point particles. (If there is significant pressure in the disc, the influence on what follows will be minor.) Consequently, their trajectories are geodesics in the space-time of the black hole. The gravity from the star and disc are assumed to be inconsequential compared to that of the hole, so the vacuum metric may be used to calculate these trajectories.

The accretion disc around the black hole is assumed to be thin and Keplerian. The outer radius is set as 20 gravitational radii ($R_g$). In this paper we use units with $G=c=1$ for simplicity,  which gives $R_g = M$ (the mass of the black hole). $S=J/M$ is the angular momentum per unit mass for the black hole, and $\tilde{S}=S/M$ is defined as the spin, which fall within the range between -1 and 1. $\tilde{S}=0$ corresponds to a spinless Schwarzschild black hole and $\tilde{S}=1$ corresponds to a maximally rotating Kerr black hole. The disc's inner radius is the innermost stable circular orbit (ISCO), which is at $9M, 6M$, and $1.237M$ respectively for black holes with $\tilde{S}=-1, 0, \textrm{ and } 0.998 $. We take spin $\tilde{S}=0.998$ instead of $\tilde{S}=1$ as a reasonable upper bound of spin (e.g., \citet{Thorne:74}).

The collision of the star and the gas transfers part of the star's kinetic energy to heat or radiation. This energy per orbit lost is small compared with the kinetic energy of the star, leaving the motion of the star roughly unaffected. We also assume we are well outside the tidal radius: $r\gg M_8^{\frac{1}{3}}$ for main sequence stars. \footnote{A magnetized neutron star provides an interesting possibility as it can present a large effective area close to the horizon of an intermediate mass hole.}

The trajectory of a star is fully determined by its initial position and momentum. Eight initial conditions are required: $t_0, r_0, \theta_0, \phi_0, \dot t_0, \dot r_0, \dot \theta_0, {\rm and}\ \dot \phi_0$.
The energy $E$, angular momentum $L$ and Carter constant $Q$ \citep{Carter} can all be derived from these initial parameters. The observer is defined to be at some inclination $\theta$ to the black-hole spin direction looking down on the accretion disc at the flares. A fifth-order Runge-Kutta numerical method is used to calculated how a particle moves given its position and momentum.

\subsection{Thin Keplerian disc}

We use the Boyer-Lindquist metric to describe the SMBH space-time:

\begin{eqnarray}
   ds^2&=&-\left(1-\frac{2Mr}{\Sigma}\right)dt^2-\frac{4SMr\sin^2\theta}{\Sigma}dtd\phi 
   + \frac{\Sigma}{\Delta}dr^2  \nonumber \\
  &&\hspace{-0.5cm}+\Sigma d\theta^2 + \left(r^2+S^2+\frac{2S^2Mr\sin^2\theta}{\Sigma}\right)\sin^2\theta d\phi^2 \ ,
\end{eqnarray}
with $\Sigma = r^2+S^2\cos^2\theta$, $\Delta= r^2-2rM+S^2$.

We assume that the accretion disc is equatorial and Keplerian. For such a disc, its gas particles will have an angular velocity:
\begin{eqnarray}
\label{orbitalf}
 \omega_k = \left(S+\sqrt{\frac{r^3}{M}} \right)^{-1}
\end{eqnarray} \citep{Misner:73} for a prograde orbit. 

Combing with the condition that the inner product of the four velocity of the gas should be -1, we have:
\begin{eqnarray}
\label{mediummotion}
   \dot{t}&=&\zeta\left(S+\sqrt{\frac{r^3}{M}}\right), \nonumber\\
   \dot{\phi}&=&\zeta\ ,
\end{eqnarray}
where 
\begin{equation}
   \zeta = \left(\frac{r^3}{M}-3r^2+2 S \sqrt{\frac{r^3}{M}}\right)^{-\frac{1}{2}} .
\end{equation}
These equations describe the four-velocity of the gas particles in the disc.

\subsection{Inferring stellar orbital elements from observations}

We treat stars as massive point particles following geodesics. The trajectory of a bound star in general relativity is approximately an ellipse, described by the orbital elements: $a_r$  (the semi-major axis of the orbit), $e$ (the eccentricity), $i$ (the orbital inclination angle with respect to the accretion disc), $\Omega$ (longitude of the ascending node N with respect to the projection of the line of sight on the disc), $\omega$ (angular distance between N and peribothon), and $T_0$ (the epoch). The three-dimensional orientation of stellar orbits varies due to two relativistic effects. Firstly, the peribothon precesses in the orbital plane. Secondly, if the black hole is spinning, the orbital plane will precess about the spin of the hole due to the dragging of the inertial frames.

Figs. \ref{a0orbit} and \ref{a0.998orbit} show the trajectories of stars with arbitrary initial parameters with black hole spins $\tilde{S}$=0 and $\tilde{S}$=0.998, respectively. We can see that in the $\tilde{S}$=0 case, the star's closest approach to the black hole precesses yet the motion remains on the same plane, forming a rosette pattern. By contrast in the $\tilde{S}$=0.998 case, the star's orbit is no longer planar - the orbital plane precesses around the spin of the black hole as well. The resulting trajectories look more like a ball of tangled yarn, with the amount of extra precession depending on the peribothon and the inclination of the stellar orbit, and the spin of the hole. For both setups, we can see that precession is rapid close to the hole, thus making the orbit quite irregular.

The mass of the hole and the size of the stellar orbit need to satisfy different conditions in order to observe general relativistic features of the stellar orbit around the SMBH: e.g. produce enough flares in reasonable observing time, reveal the eccentricity through the apsidal motion, reveal the spin through the Lense-Thirring effect, have a gravitationally stable stellar orbit, and avoid tidal disruption. If these constraints are satisfied, we may be able to derive the parameters of the system from the temporal behavior of the signals. This provides, in principle, a way to measure the spin of the black hole through accurate timing measurements.

\subsubsection{The size bound}
The ISCO describes the closest stable circular orbit around the black hole. For a black hole with mass $M$ and spin $\tilde{S}$, the semi-major radius of the orbit $a_r$ satisfies the equation \citep{Bardeen:72}:

\begin{equation}
\left(\frac{a_r}{M}\right)^2-6 \frac{a_r}{M}+8\tilde{S}\sqrt{\frac{a_r}{M}}-3\tilde{S}^2=0.
\end{equation}
Therefore, $a_r$ increases linearly with $M$ for a given $\tilde{S}$. If we choose a maximal value of $\tilde{S}=0.998$, we get the lower limit constraint on the size of a circular equatorial stellar orbit for the star not to plunge into the event horizon of the black hole. This corresponds to \citep{Misner:73}
\begin{equation}
   a_r \geq 1.237 M 
\end{equation}
for a circular stellar orbit around a spin 0.998 black hole, which in cgs unit is:
\begin{equation}
M \leq 5 \times10^{6} \left(\frac{a_r}{10^{12} {\rm cm}}\right) M_\odot.
\end{equation}

\subsubsection{The period bound}
If the period is too large, then useful observations will take too long. Therefore, we set the orbital period $P$ to be less than 1 year to obtain enough flares in a reasonable amount of time. 

Here the spin term is relatively unimportant when the star is far from the black hole. In this regime, for a circular orbit with period less than $P$, the relationship between $M$ and $a$ can be derived from equation (\ref{orbitalf}). 
In cgs units, we have:
\begin{equation}
M \geq 8 \times 10^{-6} \left(\frac{a_r}{10^{12} {\rm cm}}\right)^3 \left(\frac{P}{1 {\rm yr}}\right)^{-2} M_\odot.
\end{equation}

\subsubsection{The apsidal motion bound}
The precession rate of peribothon for an equatorial orbit is:

\begin{equation}
   \dot {\omega} =\frac{\delta\omega}{P} = \frac{M^\frac{3}{2}}{a_r^\frac{5}{2} (1-e^2) } \left(3-\frac{4S} {\left( M a_r (1-e^2)\right)^{\frac{1}{2}} } \right)
\end{equation}
(e.g., \citet{Ciu:95}).

Therefore, only for stellar orbits with large eccentricity can we observe the peribothon advance; at the same time the peribothon advance will give an estimate of the eccentricity. When the $\dot {\omega}$ is large enough to be observed, say $\dot {\omega} \geq \dot{\omega}_{\rm min}$, ignoring the small second term, we need:
\begin{equation}
   M \geq \left(\frac{(1-e^2) \dot{\omega}_{\rm min}}{3}\right)^\frac{2}{3} a_r^\frac{5}{3}.
\end{equation}
with a small correction if the black hole spins.
Taking the peribothon precession period to be shorter than $10$ years, and the eccentricity to be 0.5 for example, we have:

\begin{equation}
M  \geq  2 \times10^{2} \left(\frac{a_r}{10^{12} {\rm cm}}\right)^\frac{5}{3} M_\odot.
\end{equation}

\subsubsection{Lense-Thirring observability}
If the black hole is spinning and the orbital plane of the stellar trajectory is inclined with respect to the equatorial plane of the hole, the orbital plane will precess about the spin of the hole due to the dragging of inertial frames. This causes the precession of $\Omega$, the longitude of the ascending node. We have \citep{Ciu:95}:
\begin{equation}
  \dot {\Omega}=\frac{2\tilde{S}M^2}{a_r^3 (1-e^2)^\frac{3}{2}} 
\end{equation}

Therefore, in order to observe this $ \dot {\Omega} \geq \dot {\Omega}_{\rm min}$, 
\begin{equation}
  M \geq \left(\frac{\dot{\Omega}_{\rm min}}{2\tilde{S}}\right)^\frac{1}{2} (1-e^2)^\frac{3}{4} a_r^{3/2}.
\end{equation}
The Lense-Thirring effect can reveal the spin of the black hole.

For an orbit with $e<0.5$, a Lense-Thirring precession period less than 10 yr, and $\tilde{S}<0.998$, we will have:
\begin{equation}
  M  \geq 3 \times 10^{3} \left(\frac{a_r}{10^{12}{\rm cm}}\right)^\frac{3}{2} M_\odot.
\end{equation}

\begin{figure}
 \centering
 \includegraphics[width=3in]{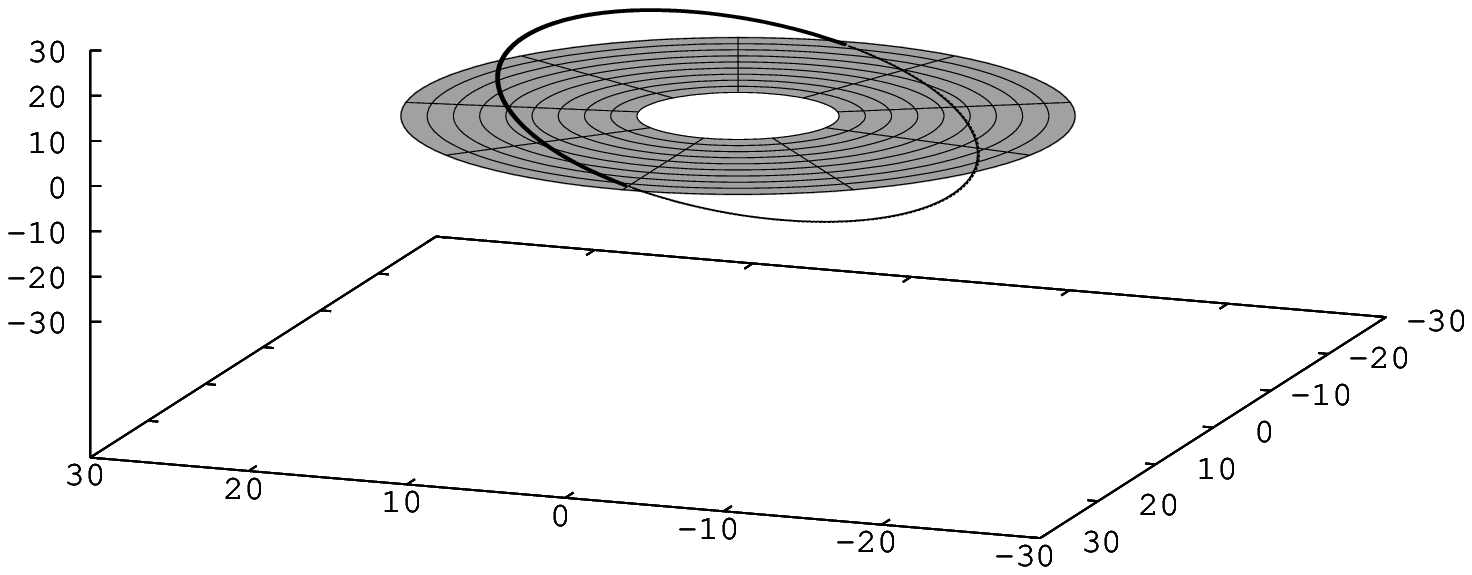}
 \includegraphics[width=3in]{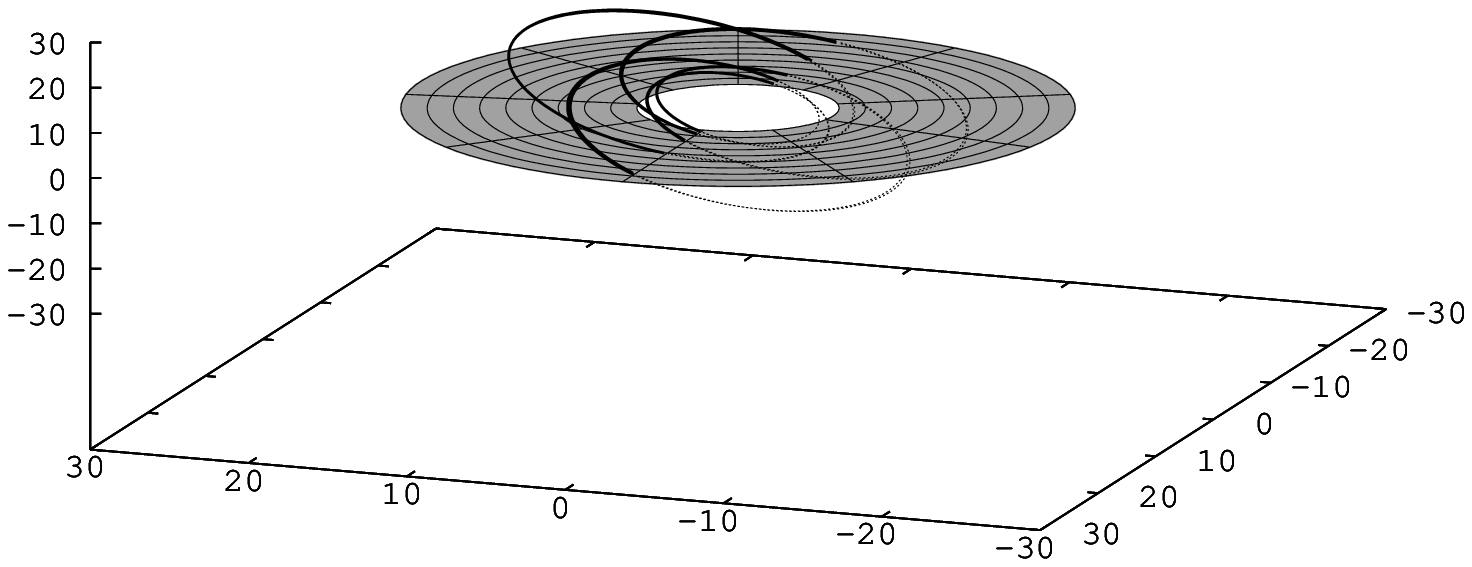}
 \includegraphics[width=3in]{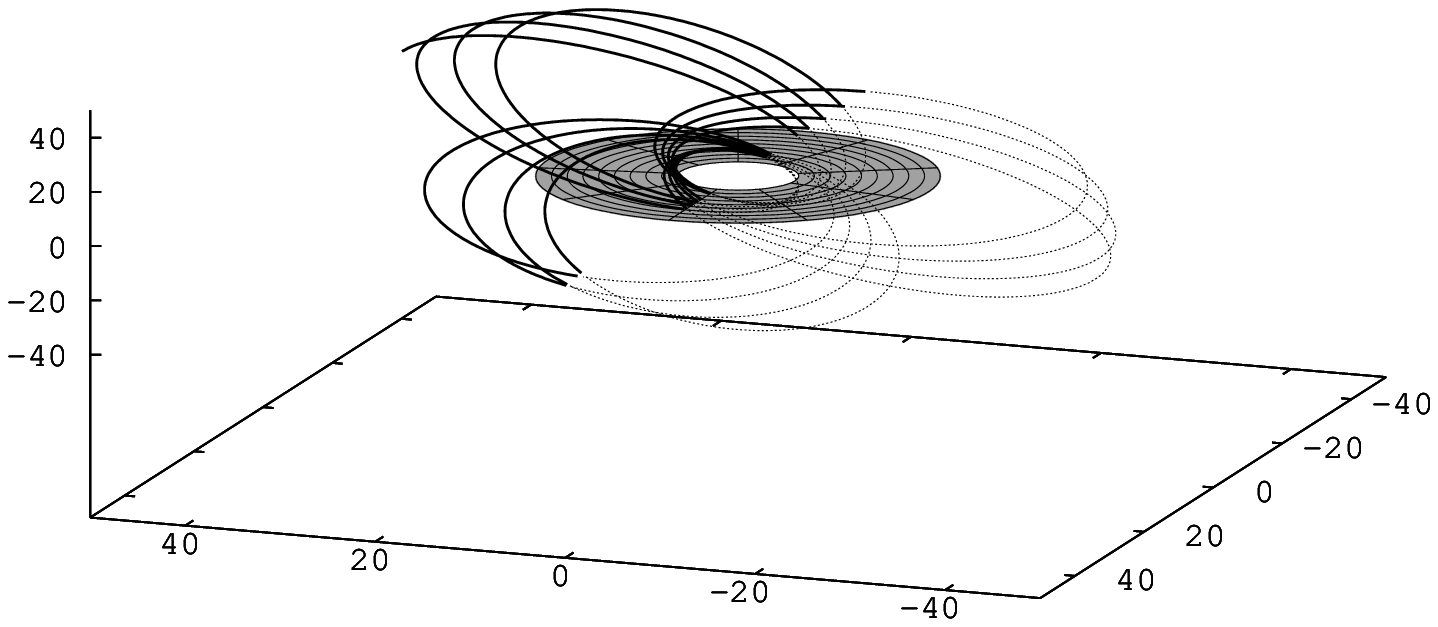}
 \includegraphics[width=3in]{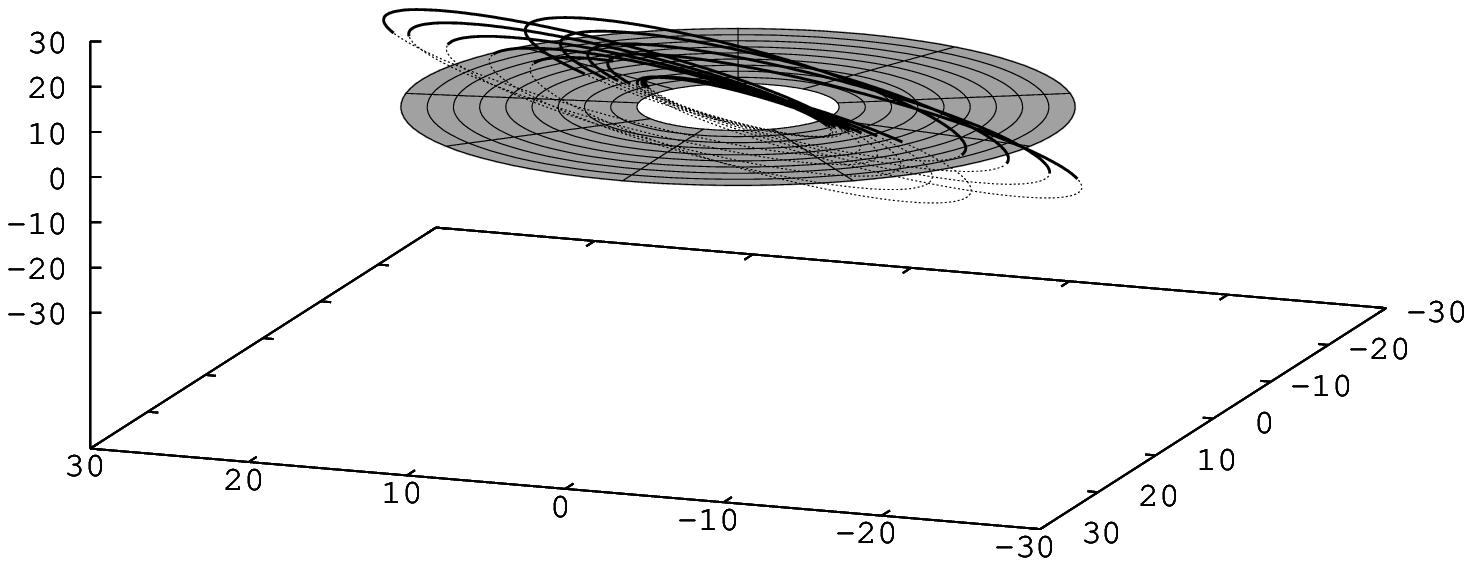}
 \caption{Possible stellar trajectories around a slowly rotating black hole: the horizontal `web' structure is the Keplerian accretion disc, ranging from 6 to 20 $M$. The four plots represent from top to bottom, stellar orbits with \textbf{a.} small eccentricity; \textbf{b.} large eccentricity; \textbf{c.} large distance from the black hole; \textbf{d.} close distance to the black hole.  Note that in each plot, the motion of the star is confined to one plane.  The orbits are denoted by thick lines above the accretion disc plane, and thin lines below.} 
 \label{a0orbit}
\end{figure} 

\begin{figure}
 \centering
 \includegraphics[width=3in]{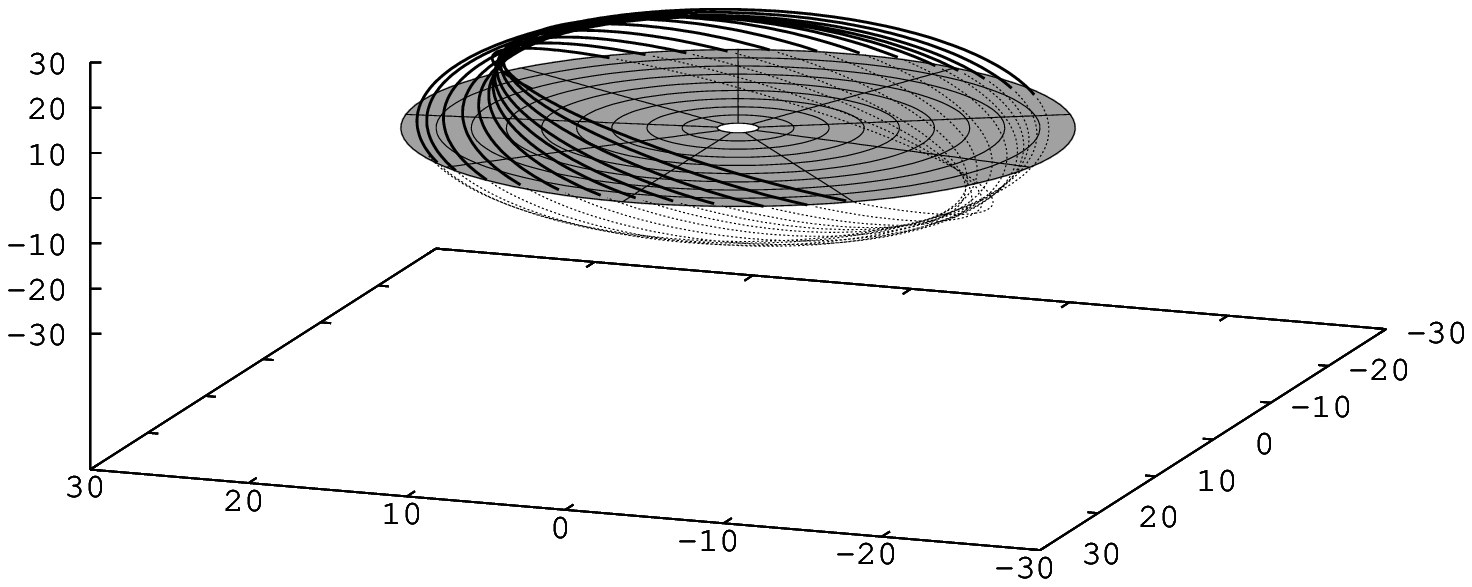}
 \includegraphics[width=3in]{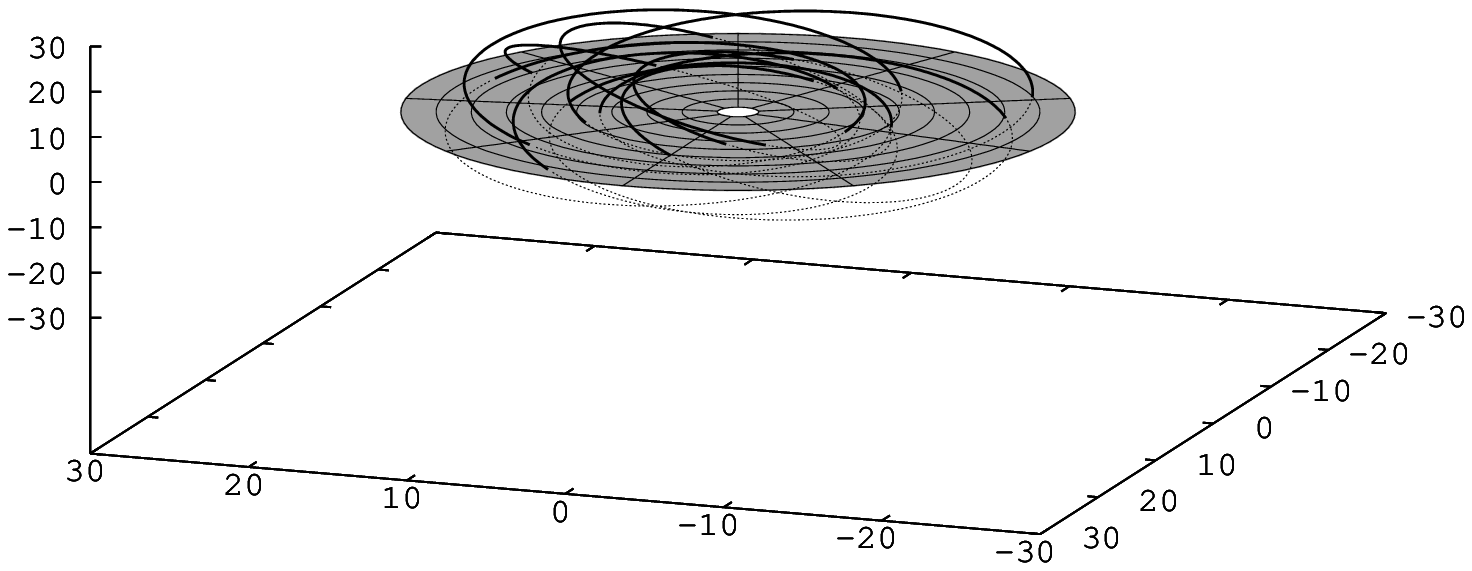}
 \includegraphics[width=3in]{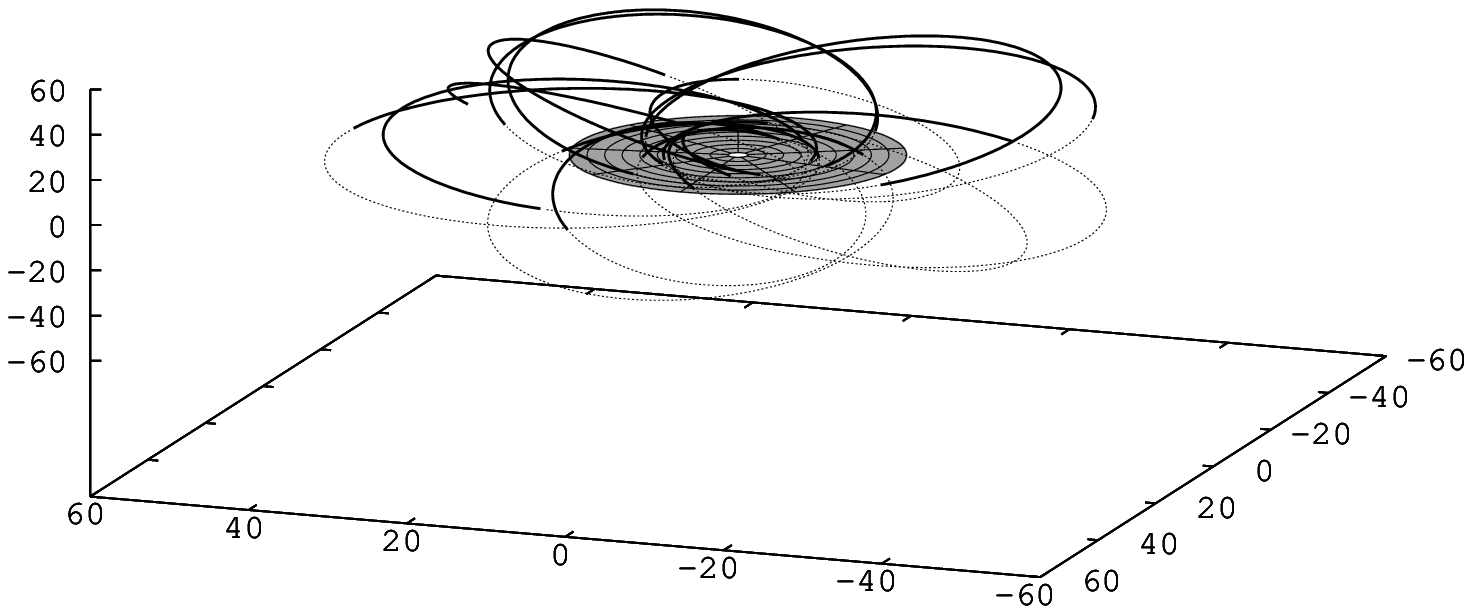}
 \includegraphics[width=3in]{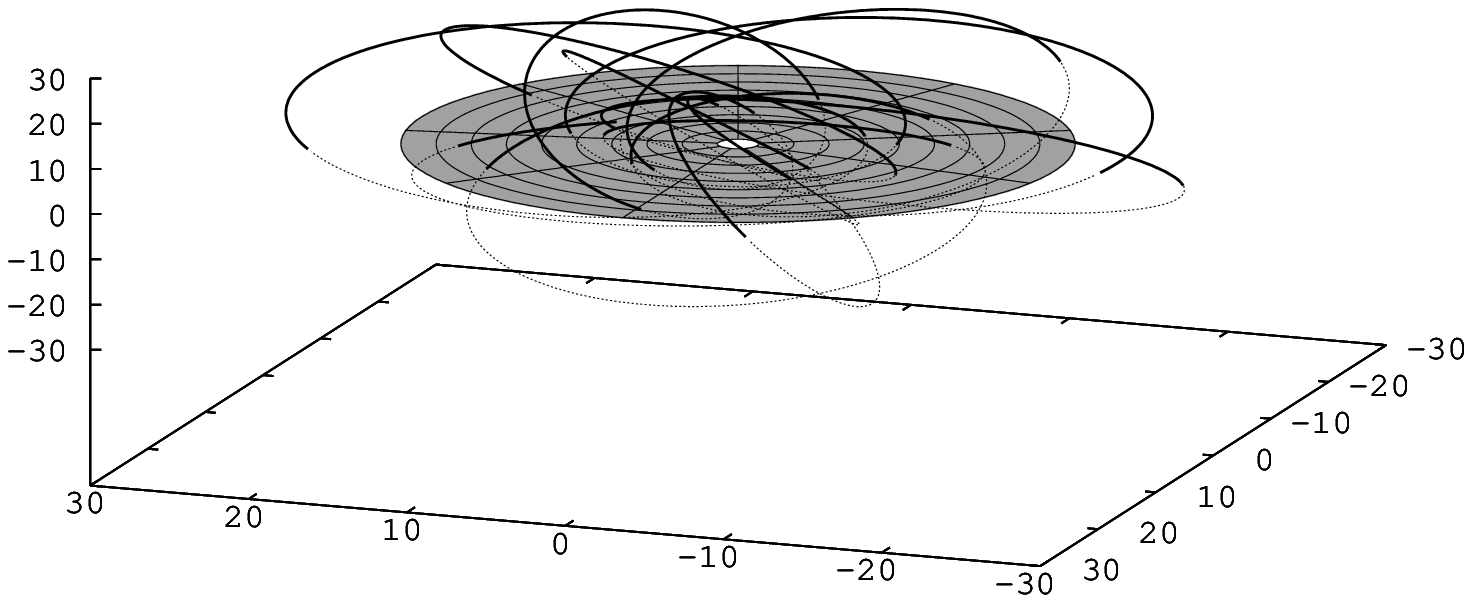}
 \caption{Possible stellar trajectories around a rapidly rotating black hole with $a=0.998$. The initial velocity and position of the star in each plot are the same as those of the corresponding stars' in Fig.~\ref{a0orbit}. Note that the motions are no longer planar due to dragging of inertial frames by the Lense-Thirring effect.} 
 \label{a0.998orbit}
\end{figure}

\clearpage

\subsubsection{Observability in the presence of gravitational radiation}
Gravitational radiation will carry off the orbital energy of the star.  Eventually the star will be drawn in towards the black hole and plunge through the event horizon or be tidally disrupted. The gravitational radiation reaction time has to be reasonably long in order for the star to be observed.

This time scale can be expressed as \citep{Misner:73}: 
\begin{equation}
  \tau = \frac{5}{256} \frac{a_r^4}{M^2 M_\star f(e)} \ \textrm{for a non-spinning hole}.  
\end{equation}
Here $M_\star (\ll M)$ denotes the mass of the orbiting star, and $f(e)=(1+\frac{73}{24}e^2 + \frac{37}{96}e^4) (1-e^2) ^{-\frac{7}{2}}$ \citep{Peters:63}.

For the stellar orbit to last for a useful finite time scale $\tau_0$, the $M-a_r$ relationship is:
\begin{equation}
  M \leq a_r^2\sqrt{\frac{5}{256 \tau_0 M_\star f(e)}}
\end{equation}
with a small correction if there is a nonzero $\tilde{S}$ or in the strong field limit. 

Suppose we have observations of $N_{\rm AGN}$ AGNs and each one swallows $N_\star$ stars over its 10 Gyr lifetime. In order to find one example, we need:
\begin{equation}
 \frac{\tau_{GR}}{10 Gyr} \geq \frac{1}{N_\star} \frac{1}{N_{\rm AGN}}. 
\end{equation}

Say $N_\star = 10^{-3} M_\star / M_\odot$ with $M_\star \leq 1M_\odot$. If we observe $10^{10}, 10^7$, and $10^4$ AGNs, the radiation reaction time is then respectively a thousand years, a million years, or a billion years, and the formula becomes:
\begin{equation}
  M \leq  4 \times10^{4}\left(\frac{a_r}{10^{12}{\rm cm}}\right)^2 M_\odot,
\end{equation}

\begin{equation}
  M \leq  10^{3}\left(\frac{a_r}{10^{12}{\rm cm}}\right)^2 M_\odot,
\end{equation}
and 
\begin{equation}
  M \leq  36 \left(\frac{a_r}{10^{12}{\rm cm}}\right)^2 M_\odot.
\end{equation}

\subsubsection{Tidal disruption bound}
The tidal radius is (e.g. \citet{Frank}):
\begin{equation}
  r_T =  5 \times 10^{12} \left(\frac{M}{10^6 M_\odot}\right)^\frac{1}{3} \frac{R_\star}{R_\odot} \left(\frac{M_\star}{M_\odot}\right)^{-\frac{1}{3}} {\rm cm}\ ,  
\end{equation}
where $R_\star$ is the stellar radius.

Therefore, for a star to avoid tidal disruption at a distance $a_r$ to the black hole, the mass of the hole has to satisfy:
\begin{equation}
  M \leq  \left(\frac {a_r}{5 \times 10^{12}{\rm cm}} \frac{R_\odot}{R_\star}\right)^3  \frac{M_\odot}{M_\star} \times 10^6 M_\odot.
\end{equation}

For a main sequence star with mass less than a solar mass would satisfy $R_\star \sim M_\star^{0.8}$ \citep{Kippenhahn}, we have:
\begin{equation}
  M \leq  \left(\frac {a_r}{5 \times 10^{12}{\rm cm}} \right)^3  \frac{M_\odot^4}{M_\star^{3.4}} \times M_6
\end{equation}
which is:
\begin{equation}
   M \leq  8 \times10^{3} \left(\frac{a_r}{10^{12}{\rm cm}}\right)^3  M_\odot
\end{equation}
for a sun-like star, 
\begin{equation}  M \leq  4 \times10^{5}  \left(\frac{a_r}{10^{12}{\rm cm}}\right)^3  M_\odot
\end{equation}
for a dwarf star (M8) (taking $R_\star = 0.13 R_\odot$, and $M_\star = 0.1 M_\odot$), and
\begin{equation}
   M \leq  7 \times10^{9} \left(\frac{a_r}{10^{12}{\rm cm}}\right)^3  M_\odot
\end{equation}
for a white dwarf (taking $R_\star = 0.009 R_\odot$, and $M_\star = 0.6 M_\odot$).

\begin{figure}
 \centering
 \includegraphics[width=3in]{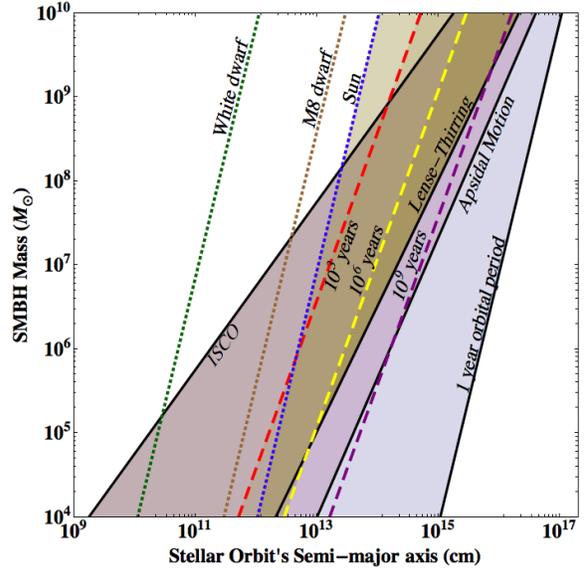}
 \caption{
 \textbf{Black solid lines} (from right to left) -
Line 1: To the left of this line, the period of a stellar orbit is less than 1 year.
Line 2: To the left of this line, the apsidal motion of the peribothon (taking arbitrary $e=0.5$) has a period less than 10 years.
Line 3: To the left of this line, the Lense-Thirring effect becomes important, say, with a period less than 10 years. $\tilde{S}=0.998$ and $e=0.5$.  
Green: To the left of this line, the star will plunge into the blackhole. $\tilde{S}$ is taken as 0.998 to plot the ISCO.
\textbf{Dashed lines}: Each line corresponds to a particular gravitational radiation reaction characteristic time scale. To the right of each line, this time scale is more than (from right to left)-Purple: $10^9$ years; Yellow: $10^6$ years; Red: $10^3$ years. 
\textbf{Dotted lines}: To the right of these lines, the star will not be tidally disrupted. Each line (from right to left) corresponds to the tidal radius for:
Blue: a sun-like star;
Brown: a M8 dwarf star with radius $0.13 R_\odot$, mass $0.1 M_\odot$;
Dark Green: a white dwarf with radius $0.009 R_\odot$, mass $0.6 M_\odot$. 
\textbf{Example}: The darkest shaded region corresponds to a stellar orbit for a sun-like star with orbital period less than 1 year, apsidal motion's and Lense-Thirring effect's period less than 10 years, and stable under gravitational radiation dissipation for more than $10^6$ years, taking $e=0.5$ and $\tilde{S}=0.998$.} 
 \label{ma}
\end{figure}

\subsubsection{Summary}

We summarize these conditions in Fig. \ref{ma}, in which the darkest shaded region indicates a stellar orbit for a sun-like star with orbital period less than 1 year, apsidal motion and Lense-Thirring effects with periods less than 10 years, and gravitational radiation lifetime more than a million years.

This plot demonstrates that, given our understanding of the distribution of stars in galactic nuclei \citep[e.g.][]{Alexander, Ghez:01}), it is quite reasonable to expect an example a sample of within $10^4$ $N_{\rm AGN}$.

\subsection{Inelastic collision of star and accretion disc}

In order to elucidate principles, we make a very simple model of a star-disc collision. Numerical simulations are needed to make a more sophisticated model. 

With the knowledge of the star and gas four-momenta at any space-time point, we can calculate how much energy is radiated following the collision. We treat the star-disc collision ($v_\star, v_{gas} \gg v_{sound}$) as a totally inelastic collision of two massive particles, one with the mass of the star and the other calculated by multiplying the surface density of the disc by the collision area. We assume that the intensity of the flare is proportional to the energy deposited on the star. Here we assume that the star goes through the accretion disc, `cuts' a hole on that plane and carries away that part of the gas with it. The shape of the `hole' is an ellipse of area $A$.

Suppose that, before the collision, the four-momentum of the star is $p_{1s}^{\alpha}$ and the four-momentum of the gas is $p_{1g}^{\alpha} $. After the collision, the four-momentum of the star is $p_{2s}^{\alpha} $, the four-momentum of the gas is $p_{2g}^{\alpha} $, and the total energy and average four-velocity of the photons are $E_{\rmn{0}}$ and $u_{\gamma}^{\alpha}$. In the center of momentum frame of the photon, we have the conservation of energy-momentum of the system:

\begin{equation}
   p_{1s}^{\alpha} +p_{1g}^{\alpha} = p_{2s}^{\alpha} +p_{2g}^{\alpha}+E_{\rmn{0}}u_{\gamma}^{\alpha},
\end{equation}
which gives, 
\begin{equation}
   m_s u_{1s}^{\alpha} +m_g u_{1g}^{\alpha} = m_s u_{2s}^{\alpha} +m_g u_{2g}^{\alpha}+E_{\rmn{0}}u_{\gamma}^{\alpha}.
\end{equation}
Here $m_s$ is the mass of the star, and $m_g$ is the mass of the gas which collides with the star. The $u$'s are the corresponding four-velocities.

Just after the collision, the star and gas move with the same four-velocity, so $u_{2s}^{\alpha} = u_{2g}^{\alpha}$. In the frame of the star after collision, $u_{2s}^{\alpha}$ and $u_{2g}^{\alpha}$ are (1, 0, 0, 0). We assume that the radiation is isotropic in this frame, so $u_{\gamma}^{\alpha}$  also is (1, 0, 0, 0) since isotropic photons will have no net mome
To view a message, click on it.ntum. Squaring on both sides of the equation, we obtain

\begin{equation}
   2 m_g m_s u_{1s}^{\alpha}  u_{1g\alpha} = -E_{\rmn{0}}^2-2 m_s E_cc - 2 m_g E_{\rmn{0}} -2 m_g m_s.
\end{equation}

We assume that the radiated energy and the rest mass of the gas are much smaller than the rest mass of the star, so we ignore terms like ${E_{\rmn{0}}}^2$ and $m_g E_{\rmn{0}}$ compared to terms containing $m_s$. The remaining terms become:

\begin{equation}
   m_g m_s u_{1s}^{\alpha}  u_{1g\alpha} \simeq - m_s E_{\rmn{0}} - m_g m_s.
\end{equation}
Therefore, the radiated energy is
\begin{equation}
   E_{\rmn{0}} \simeq m_g (-u_{1s}^{\alpha} u_{1g\alpha}-1). 
\end{equation} 

Here $m_g=\sigma A$ is the mass of gas taken away by the star, where $\sigma$ is the surface density of the accretion disc, which is assumed not to change from collision to collision. Thus, the problem of calculating the radiated energy is transformed to calculating the size of the collision area in the disc rest frame.

Here we assume that space-time is flat locally. The shape of the star in its rest frame is a sphere in 3-d space, ignoring the effect of gravity exerted by the central black hole and rotation. In flat space-time, the major and minor axes of this ellipse are simply $R_\star$ and $R_\star/{\cos(\theta_g)}$, where $\theta_g$ is the angle between the normal vector of the plane and the velocity of the star in the rest frame of the gas in the disc.

Define the frame of the gas such that the accretion disc is in the xy-plane, and the normal vector is the z-axis. Suppose that $a$ and $b$ are the major and minor axes of the impact ellipse. In the gas's initial rest frame, its area is:

\begin{equation}
   A = \frac{\pi R_\star^2}{\cos\theta_g}
\end{equation}  

In the Boyer-Lindquist frame, there is a Lorentz contraction along the direction of the motion of the gas in the plane of the disc, so the area becomes
\begin{equation}
   A = \frac{\pi R_\star^2}{\cos\theta_g \gamma_g}
\end{equation}  
where $\gamma_g$ is the Lorentz factor of the gas.

Clarifying the notation: in the gas frame, the four-velocities of the gas and the star are \begin{eqnarray}
   u_g^\alpha &=& (1, 0, 0, 0),\\
   u_s^\alpha &=& \gamma_{sg} (1, v_{sg}^x,  v_{sg}^y, v_{sg}^z).
\end{eqnarray}  
Here the $x_{ab}$ means object a's quantity $x$, in b's frame, and from these two equations we have
\begin{eqnarray}
   \gamma_{sg} &=& -u_g^\alpha u_{s\alpha}, \\
   \cos\theta_g &=& |v_{sg}^z|/v_{sg}, \\
   v_{sg} &=& \sqrt{1-\gamma_{sg}^{-2}} ,
\end{eqnarray}

In the Boyer-Lindquist frame, the disc lies in the $r-\phi$ plane, and the four velocities and the unit normal vector of the disc plane are
\begin{eqnarray}
   u_g^\alpha &=& \gamma_g (1, v_g^x, v_g^z, v_g^y),\\
   u_s^\alpha &=& \gamma_s (1, v_s^x, v_s^z, v_s^y), \\
   n^\alpha &=& \Sigma^{-\frac{1}{2}}(0, 0, 1, 0).
\end{eqnarray} 

Using special relativity, we have
\begin{equation}
   v_{sg}^z = \frac{v_s^z}{\gamma_g (1- \vec v_g \cdot \vec v_s)}\ ,
\end{equation}
Here $\vec{v}_g$ and $\vec{v}_s$ are three-vectors, and the dot is an inner product. $v_s^z$ is the $\theta$ component of $\vec{v}_s$ in the Boyer-Lindquist frame, so
\begin{equation}
   v_s^z = \frac{|u_s^\alpha n_\alpha|}{\gamma_s}\ .
\end{equation} 
Also, the inner product of $u_g$ and $u_s$ gives
\begin{equation}
   1- \vec v_g \cdot \vec v_s = - \frac{u_g^\alpha u_{s\alpha}}{\gamma_g \gamma_s}.
\end{equation} 
Therefore,
\begin{equation}
   \cos\theta_g = -\frac{|u_s^\alpha n_\alpha|}{(u_g^\alpha u_{s\alpha}) \sqrt{1-\gamma_{sg}^{-2}}}.
\end{equation}

The collision area calculated in terms of four-vector scalar products is then:
\begin{equation}
   A = \pi R_\star^2 \frac{\sqrt{(u_g^\alpha u_{s\alpha})^2-1}}{\gamma_g |u_s^\alpha n_\alpha|}\ ,
\end{equation}  
yielding the total collision energy as
\begin{equation}
   E_{\rmn{0}} = -\sigma \pi R_\star^2 \frac{u_g^\alpha u_{s\alpha}+1}{\gamma_g |u_s^\alpha n_\alpha|}{\sqrt{(u_g^\alpha u_{s\alpha})^2-1}}\ .
   \label{flareenergy}
\end{equation}  

\subsection{Ray-tracing}

The energies of the photons in the observer's frame are shifted by a redshift factor $g=E/E_{\rmn{0}}$ \citep{Cunningham:75} compared to those in the emission frame. $E$ is the energy in the observer's frame and $E_{\rmn{0}}$ in the star's inertial rest frame. Noting that $E = -p^\alpha u_{\alpha}$ and $E_{\rmn{0}}=-p_{\rmn{t}}$, we have $g=p_{\rmn{t}} /(p^\alpha u_{\alpha})$, where $u_{\alpha}$ is the four-velocity of the star. Using conservation of phase-space, it is possible to show that this causes an extra intensity shift proportional to the fourth power of energy. \citet{Lindquist} shows that bolometric intensity $I=g^4 I_{\rmn{0}}$. For spectral intensity, $I_\nu = g^3 I_{\nu 0}$ at the transformed frequency.

We ray trace the photons backwards in time from the observer to the disc. Each photon has a different propagation time to the observer, due to their different emission sites on the disc, their initial momentum, and the Shapiro time delay. The different time delays of the photons contribute to the quasi-periodity of the flares. By collating the mapping between the sky-plane and the surface of the disc, it is possible to obtain the light delay time and redshift factor for each image pixel. Using this information, one can calculate the flare pattern for any stellar trajectory. This is efficiently done by sorting the pixels by light delay time and scanning only those pixels known to be within an impact region.

We use a top-hat function for the shape of the flare in time. By using such a simple functional form, we seek to describe the timing accurately. However, this would have to be convolved with a more realistic flare profile to produce a light curve to fit observational results.



\section{Results: Arrival time of the Flares}

With the knowledge of the stellar trajectory, the collision model, and the ray trajectories, we calculated the arrival time profile using the mapping between flare locations and the observer. The signals exhibit quasi-periodicity for two reasons. Firstly, due to the precession of the stellar orbit, the time separations between two consecutive collisions are different. Secondly, because the collisions happen at different locations each time, the light travel times for the signals differ. The nature of this quasi-periodicity depends on the mass and spin of the SMBH and the orbital elements of the stellar orbit, as summarized in Table \ref{table1}. Some example plots are shown in Figs. \ref{size} and \ref{spineffect}.

Eight parameters excluing the epoch $T_0$ are needed to characterize the time series in general: the mass of the black hole, the spin of the black hole, the observer's inclination angle w.r.t. the spin, and the orbital elements of the the star. In addition, two of the angle parameters are degenerate due to the symmetry of the model to rotations about the axis in the direction to the observer.

Finally, in highly symmetric orbits, other parameters may be degenerate. The eccentricity is related to the apsidal motion, and can be determined from the changing rate of the period. These two obviously become degenerate for circular orbits. The spin and the inclination are coupled by the Lense-Thirring effect, and if the black hole does not spin, the orbital inclination is not measurable.

If the peribothon is midway between two hits, the energy released in successive collisions will be similar. However, due to our assumption of isotropic emission in the star's frame, if the star is approaching the observer in a collision, the observed collision will be brighter due to Doppler boosting. These effects mean that flares will vary in brightness. Unfortunately, our model of flare flux is quite primitive, but this does not affect the use of the flare timing in determining system parameters as long as the flares can be individually distinguished.

Finally, note that the small spikes at the base of the large spikes most noticeable in Fig.~\ref{size}a are flux from higher-order images due to photons partially orbiting the black hole whilst propagating to the observer. This can provide extra information about the inner extent of the accretion disc due to it possibly blocking this flux.

\begin{table}
 \caption{Behaviors of quasi-periodicity according to the parameters}
 \label{table1}
 \begin{tabular}{@{}lcccccc}
  \hline
  Parameter & Effects on quasi-periodicity\\
  \hline

  $a_r$  &The larger the semi-major axis of the stellar
  &&\\
  & orbit, the more periodic the signals are.\\

  \hline
   $e$& A large $e$ causes the interval between 
   && \\
   & two consecutive events to differ.\\   
     
  \hline
  $i$   & Orbits with higher inclination are 
  &&\\
  & affected more by the black hole spin.\\   
  
  \hline
  $\Omega$   &  Different Roemer time delay 
  &&\\
  & and Shapiro time delay   \\
  
  \hline
  $\omega$  & Ratio of the height of flares \\

  \hline
  $\theta$ & Large observation inclination angle
  &&\\
  & increases flux intensity variability.\\  
  
  \hline
  $M$   &	Changes overall  time scale \\
  
  \hline
  $\tilde{S}$   &	Causes the time separation  
  &&\\
  & between signals to vary\\
  
 \hline
 \end{tabular}

\end{table}

\subsection{Dominant parameters}

$a_r, M, e$, and $\Omega$ basically determine the temporal behavior of the signals while the other parameters give small corrections to the quasi-periodicity or change the flux variability. Therefore, we investigate these three parameters in a Schwarzschild case since the spin is not that important here.

\subsubsection{Size and eccentricity}

\begin{figure}
 \centering
 \includegraphics[width=3in]{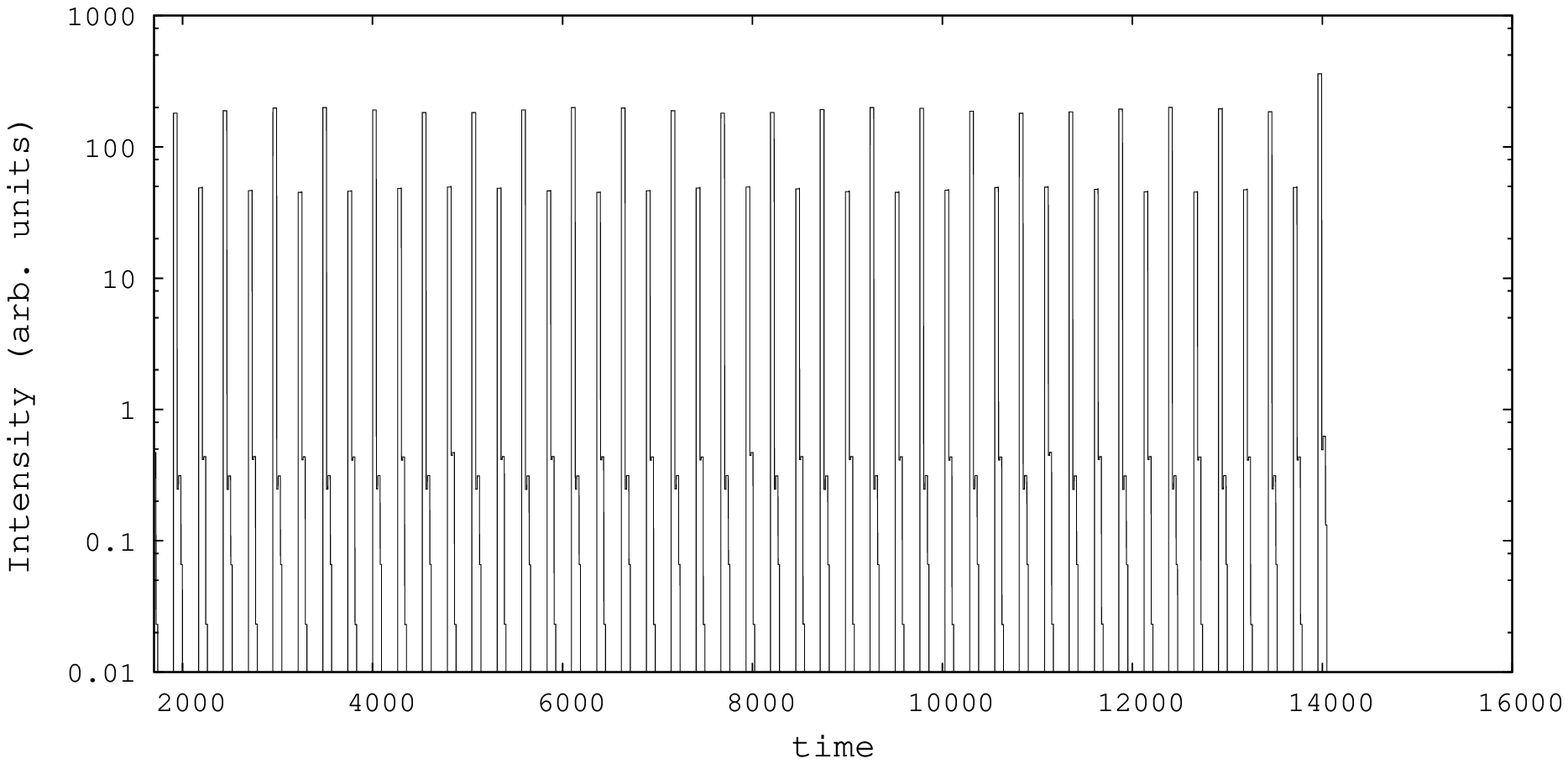}
 \includegraphics[width=3in]{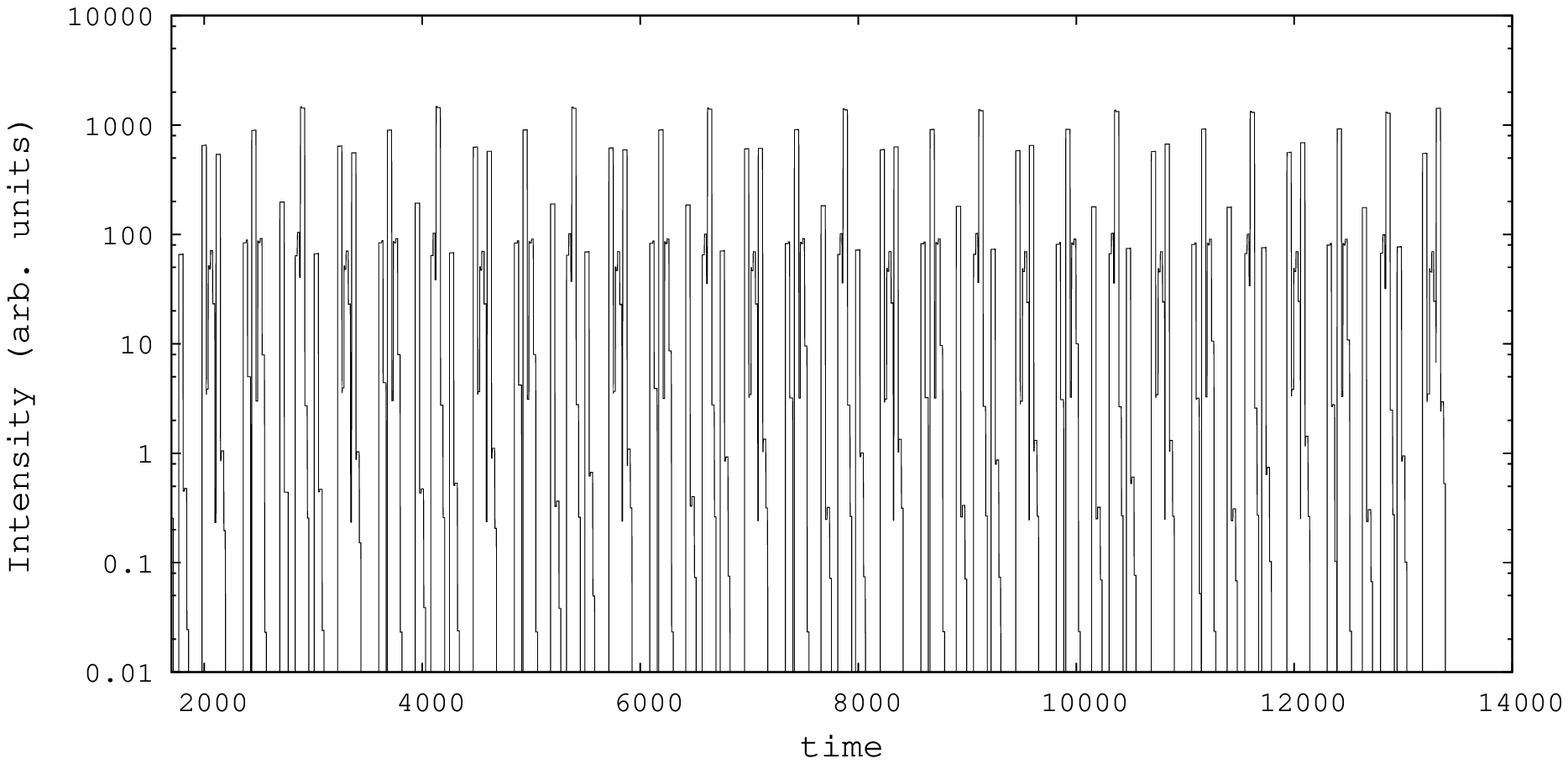} 
 \includegraphics[width=3in]{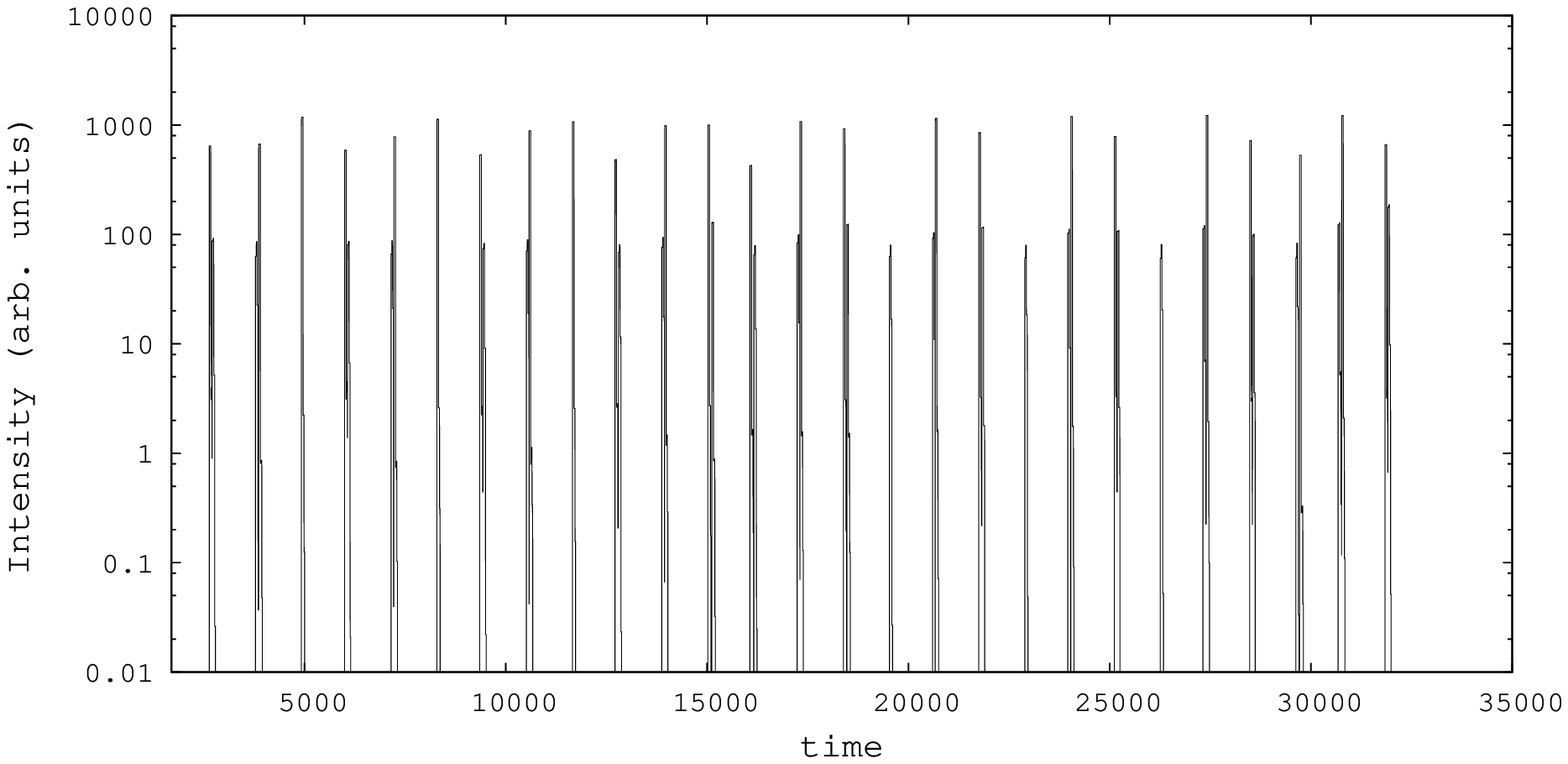}
 \includegraphics[width=3in]{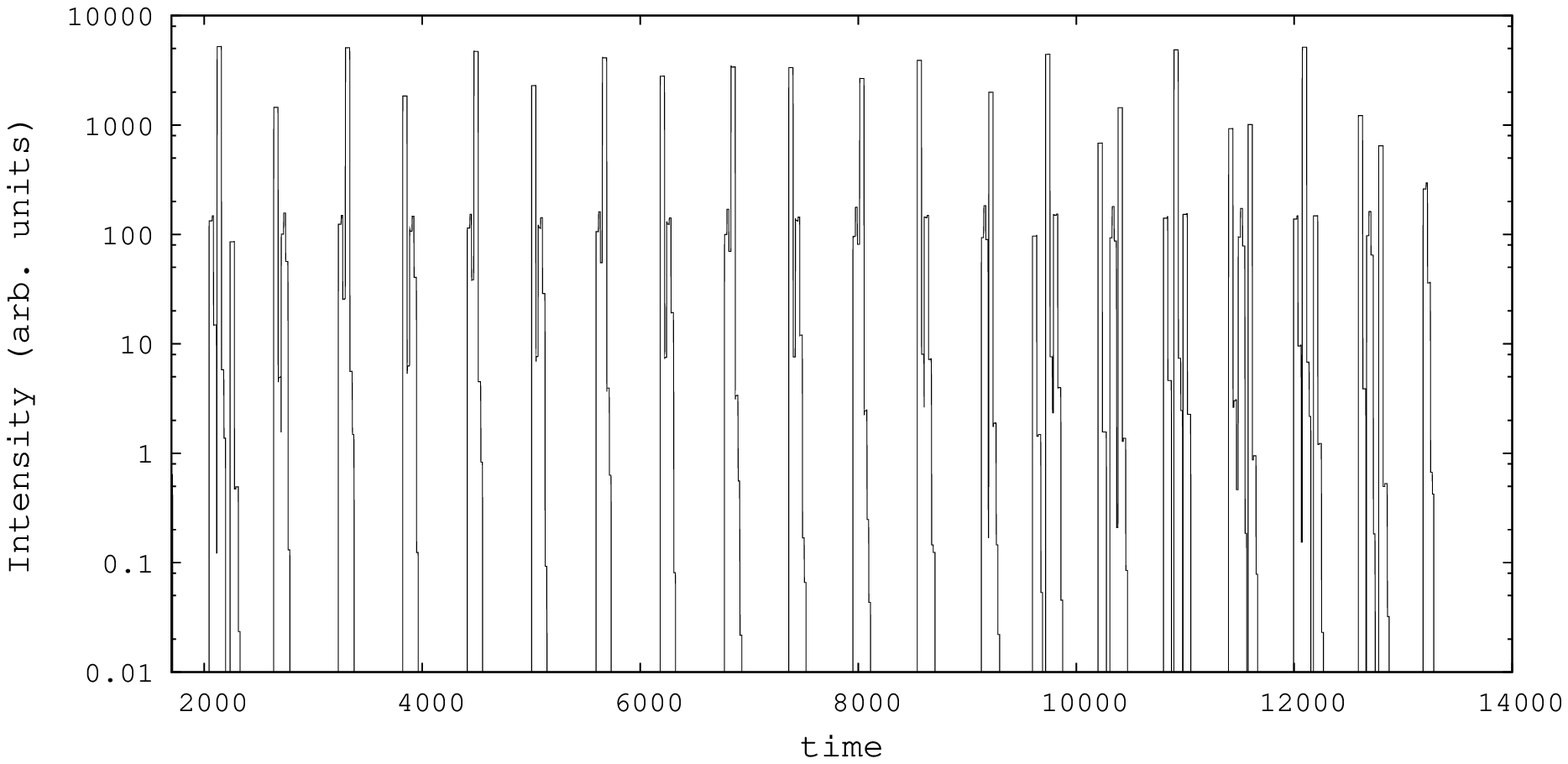}
 \caption{These are the fluxes from the flares corresponding to the stellar trajectories in Fig. \ref{a0orbit}. The x-axis is time in the units of $\frac{GM}{c^3}$. The y-axis is the fluxes of the flares in arbitrary units.  From top to bottom, in \ref{a0orbit}a, the orbit is basically circular, and consecutive flares are separated by similar time intervals. In \ref{a0orbit}b, the orbit is more eccentric, and flares are unequally spaced. The flares from \ref{a0orbit}c are much more periodic than from \ref{a0orbit}d due to the star spending more time away from the black hole.  The bottom case shows how the secular effects of precession can add up to large changes in the flare pattern.}
 \label{size}
\end{figure}

Figs. \ref{size}b and \ref{size}c show the observed flares resulting from stellar orbits with smaller and larger $a_r$ respectively, corresponding to orbits in Figs. \ref{a0orbit}b and \ref{a0orbit}c.  As can be seen, when the size of the star's orbit becomes large compared to the gravitational radius $R_g$, the signals becomes more periodic. This is because when the size of the star's orbit is large, or on average the star is far away from the black hole, general relativity plays a rather unimportant role. The star basically moves as on a Newtonian orbit, giving periodic patterns of the signals as the precessions is small. Note that the mass of the black hole only appears as a scale in the problem, via the size of $R_g$.

\begin{figure}
 \centering
 \includegraphics[width=3in]{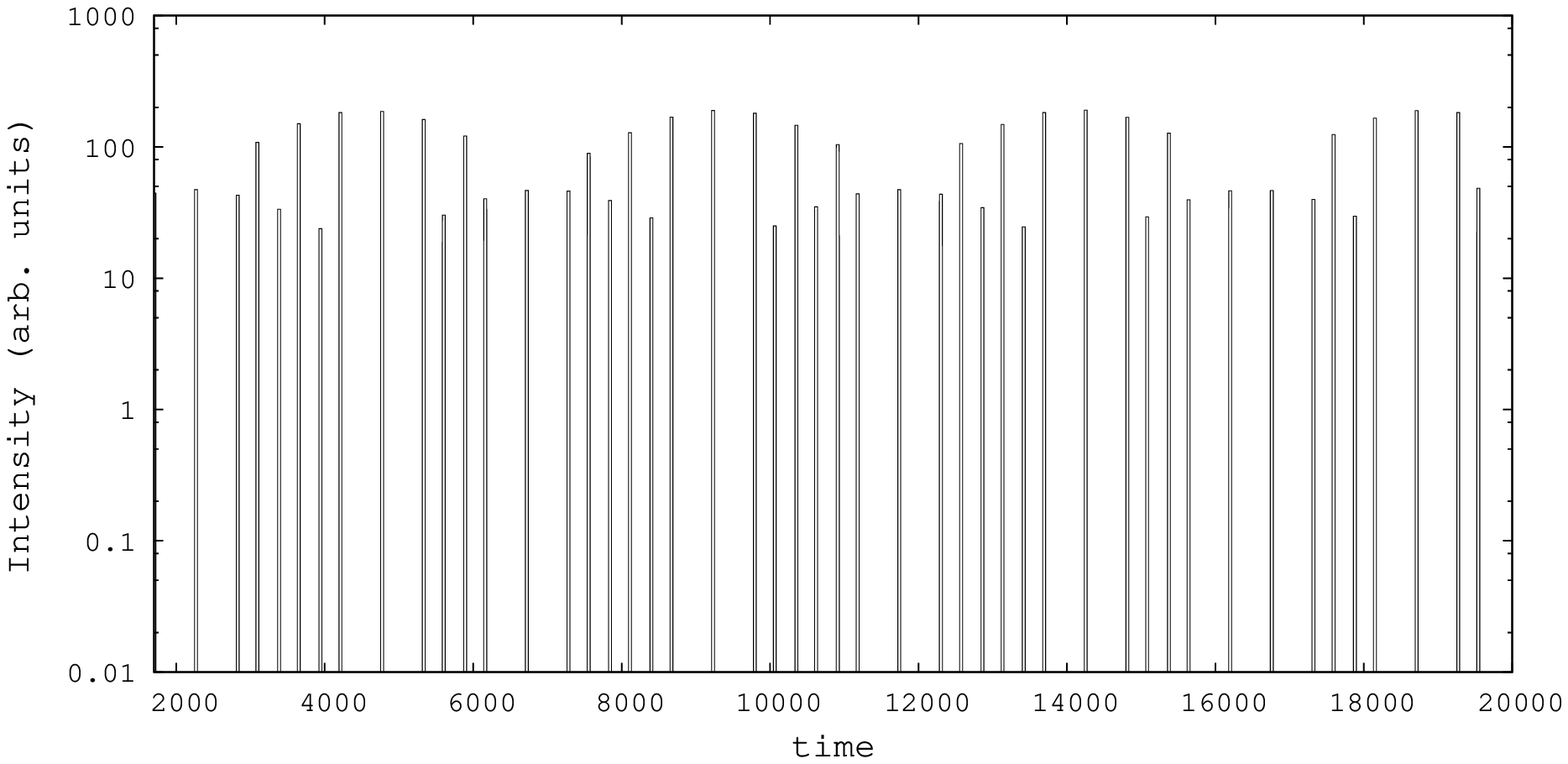}
 \includegraphics[width=3in]{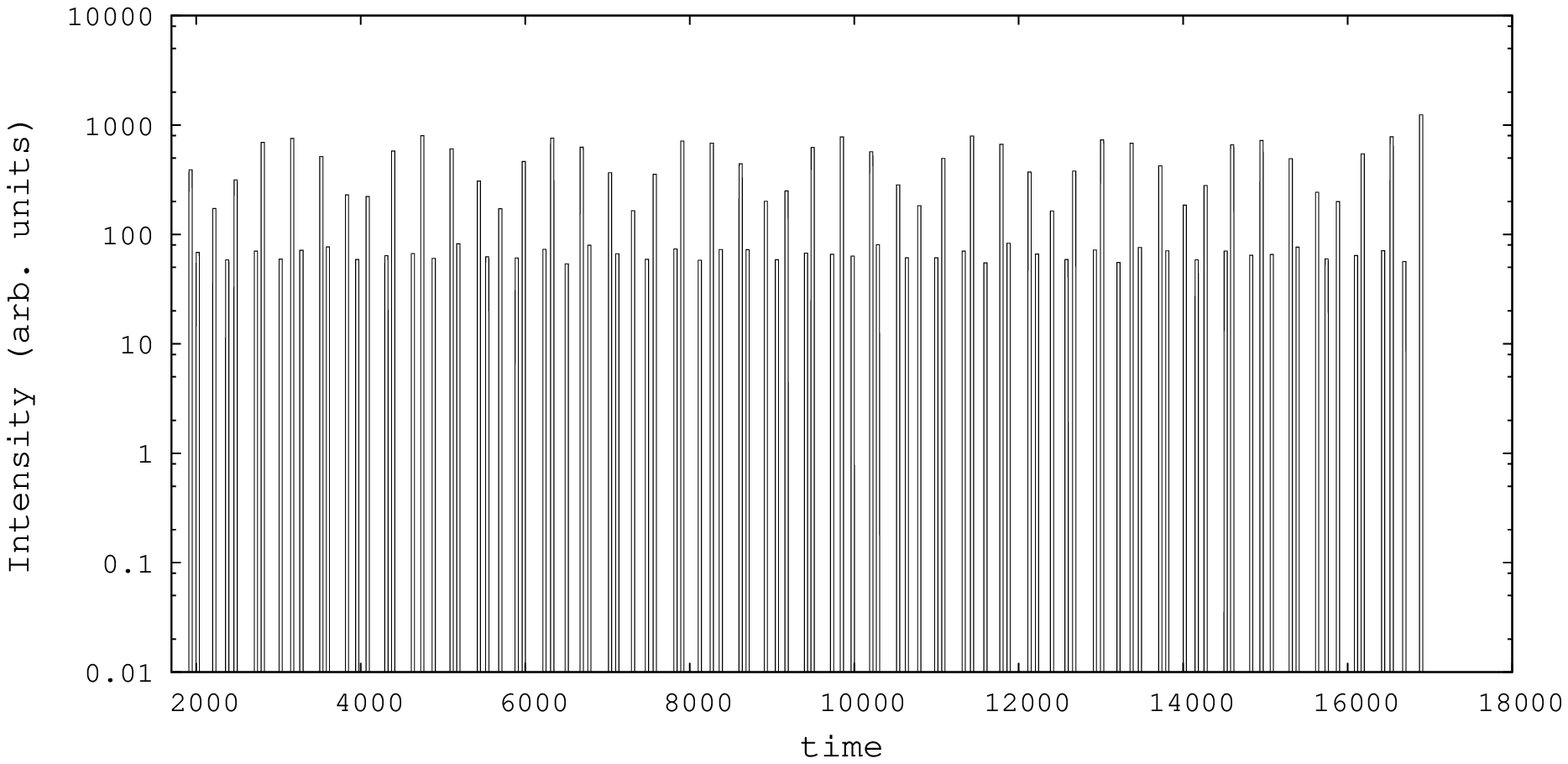}
 \includegraphics[width=3in]{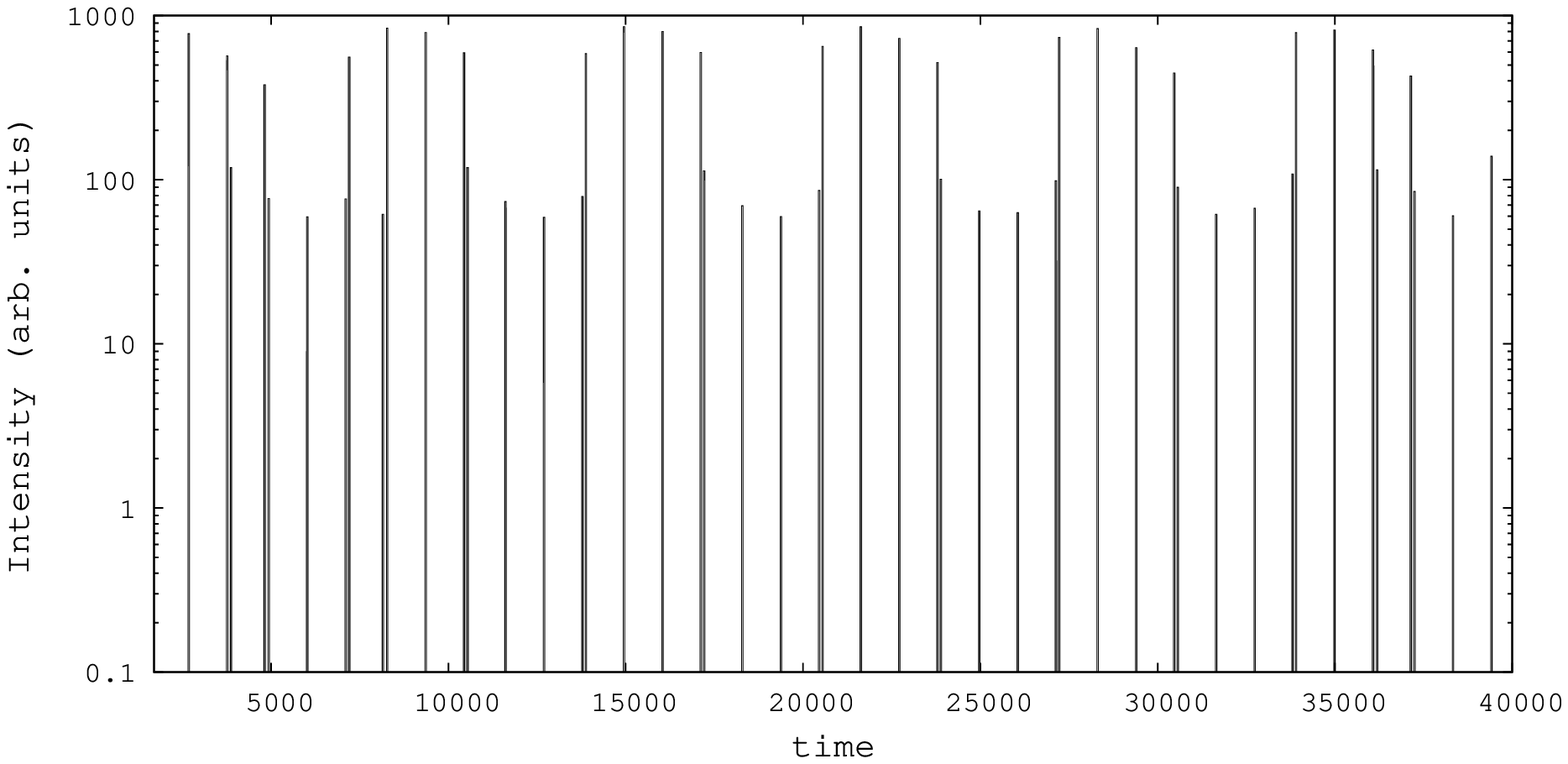}
 \includegraphics[width=3in]{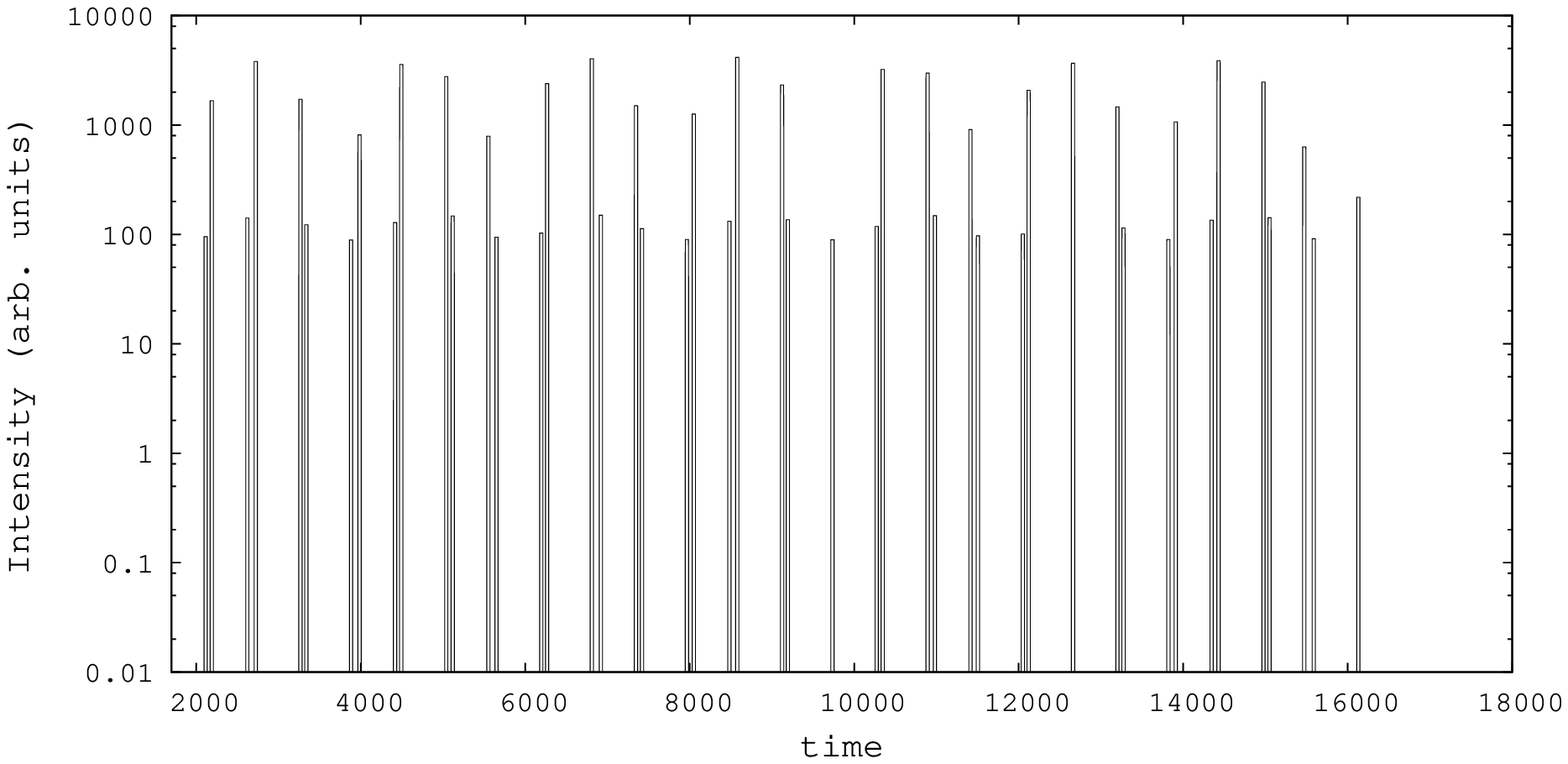}
 \caption{These are the fluxes of the flares in Fig. \ref{a0.998orbit}. The spin of the black hole affects the orbit via inertial frame dragging.  This introduces extra precession which can be seen as modulation of the flaring pattern. In principle, this allows the determination of the black hole spin from observations of star-disc collisions.}
 \label{spineffect}
\end{figure}

\clearpage

Figs. \ref{size}a and \ref{size}b show the flares from the two trajectories in Figs. \ref{a0orbit}a (low $e$) and \ref{a0orbit}b (high $e$).

\begin{figure}
 \centering
 \includegraphics[width=3in]{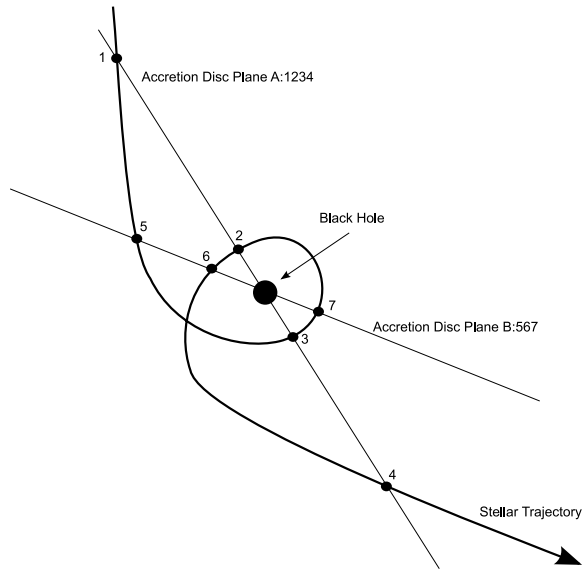}
 \caption{This figure shows how multiple flares can be produced. The star's trajectory can be strongly affected by the black hole's gravity.  This can cause more than two crossings of the accretion disc plane per close approach. Depending on the inclination of the orbit relative to the accretion disc, the extent of the disc, and the amount of doppler boosting, a varying number of closely separated flares may be seen;  i.e., for the case where the flares are produced at locations 1, 2, 3, 4;,there may be two bright, and two dim flares per close approach. For the cases 5, 6, 7; there may be either two bright and one dim, or two dim and one bright flares.  Also note that extreme gravitational lensing may produce higher-order images of flares.  This may introduce more features in the light curve if the orientiation is such that the accretion disc does not obscure them.} 
 \label{triplepeak}
\end{figure}

Consecutive signals may be grouped in multiple peaks.  The class of trajectory that does this is described in Fig. \ref{triplepeak}.  As can be seen, the time intervals between a group of three consecutive peaks can be much shorter than that between the third and fourth signals, and this pattern will repeat itself. This gives a four-peak feature, as shown in Fig. \ref{size}b. Note that if the star goes between the hole's event horizon and the inner radius of the accretion disc, or beyond the outer radius of the accretion disc (as in Fig. \ref{size}c), we will miss some hits and may observe only double peaks.

\subsubsection{Longitude of the ascending node}

To examine the role that $\Omega$ plays, we use the same stellar trajectory (as in Fig. \ref{a0orbit}b) and rotate it about the z-axis. For the original orientation, the hits' longitudes are about $0$ and $\pi$, namely, in front of and behind the black hole as seen by an observer. Then for a second orientation which is a $\pi /2$ rotation of the orginal one, the star goes through the disc on the sides of the black hole with $\Omega = \pi /2$ and ${3\pi}/2$. And for a third orientation which is a $\pi /4$ rotation, the situation is half-way between these extremes.

\begin{figure}
 \centering
 \includegraphics[width=3in]{pics/traj/hit_13_5.ps}
 \includegraphics[width=3in]{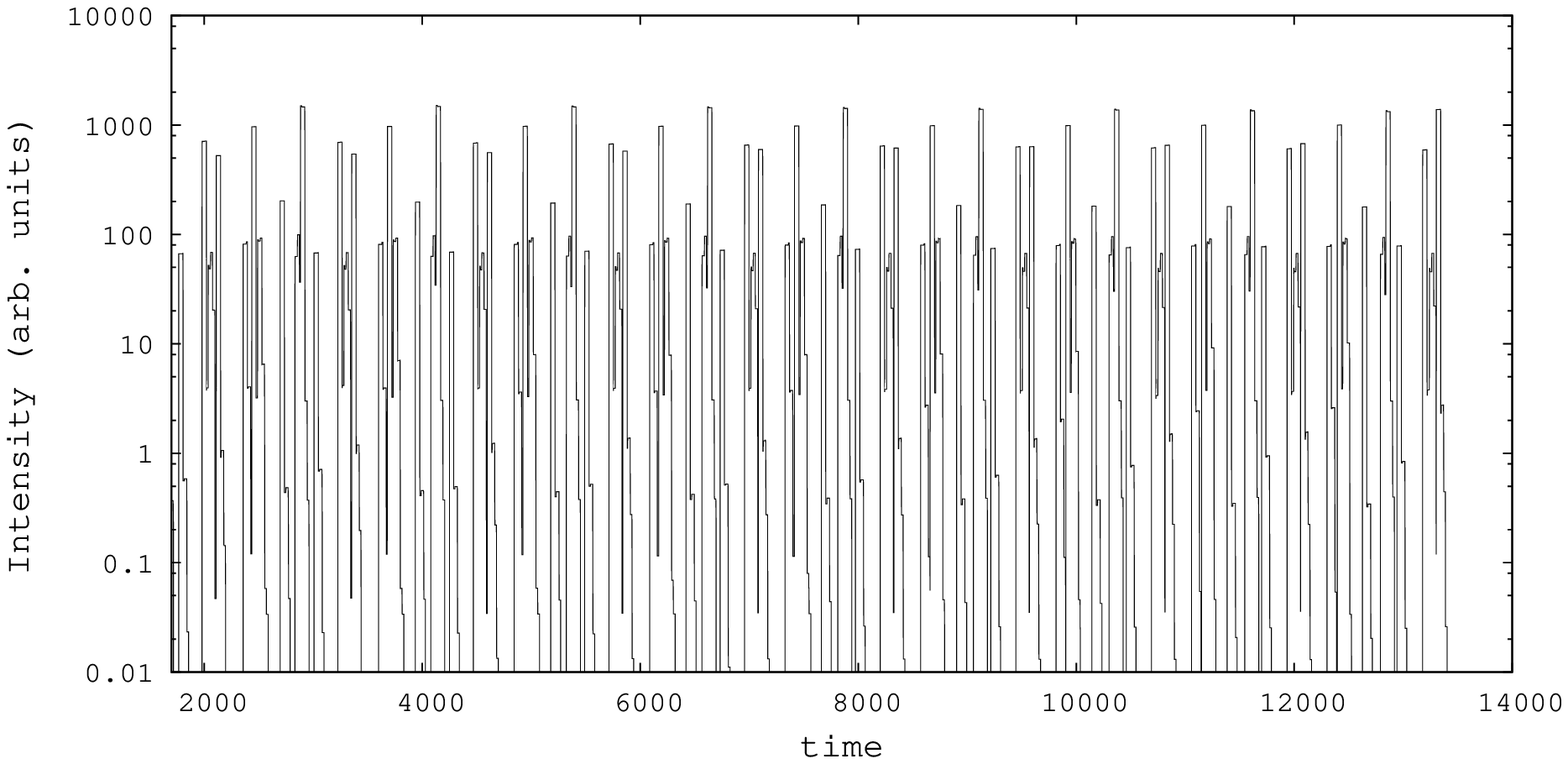}
 \includegraphics[width=3in]{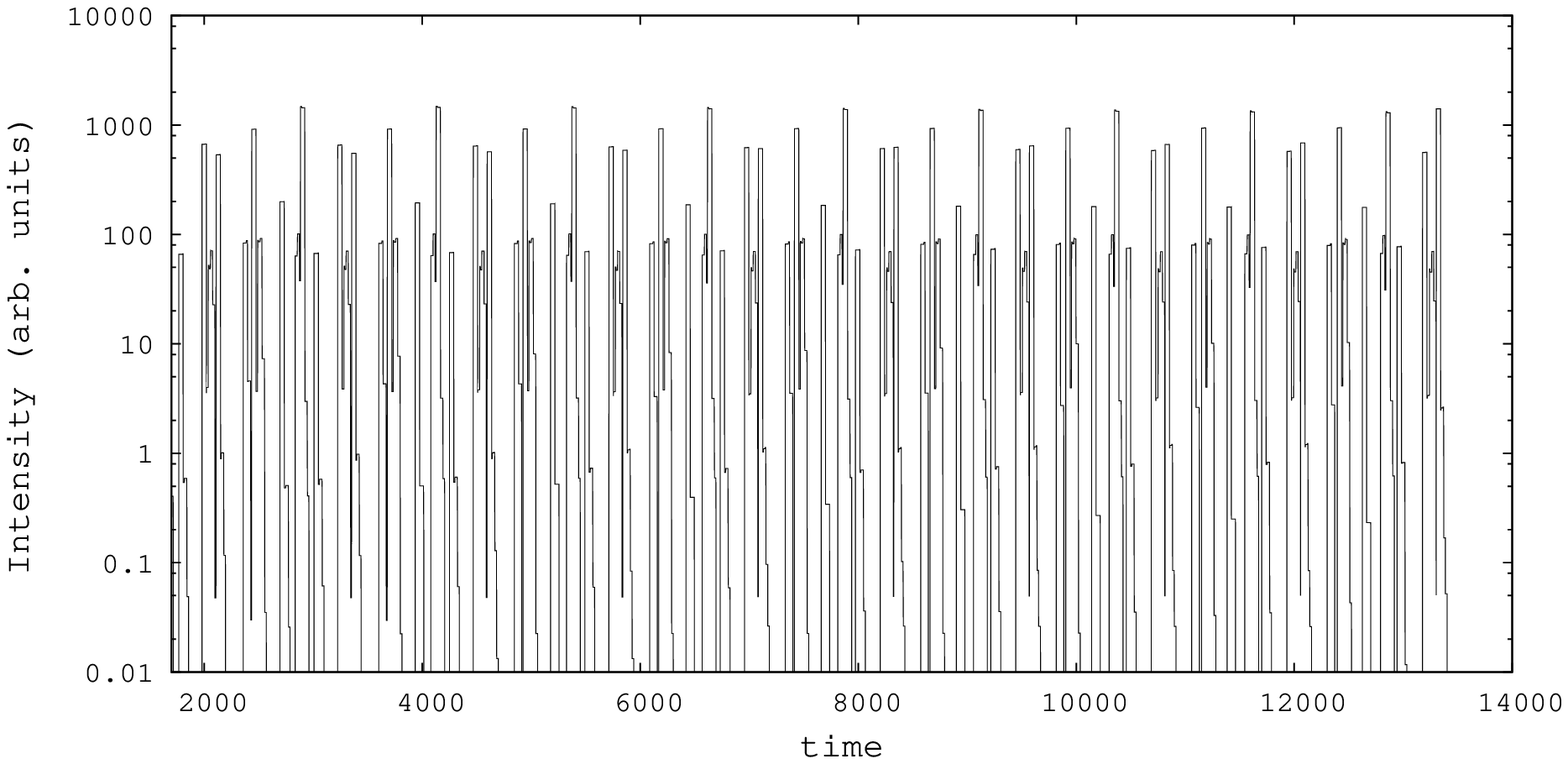}
 \caption{These fluxes are for the orbits in Fig. \ref{orbitlongitude}, when the observer is observing almost from top down at a $5^{\circ}$ angle with the spin axis of the black hole. The time intervals of any double-peak is roughly the same in any figure. The spin of the black hole is taken to be 0 for simplicity in this figure and the next one.} 
 \label{longitude5}
\end{figure}

\begin{figure}
 \centering
 \includegraphics[width=3in]{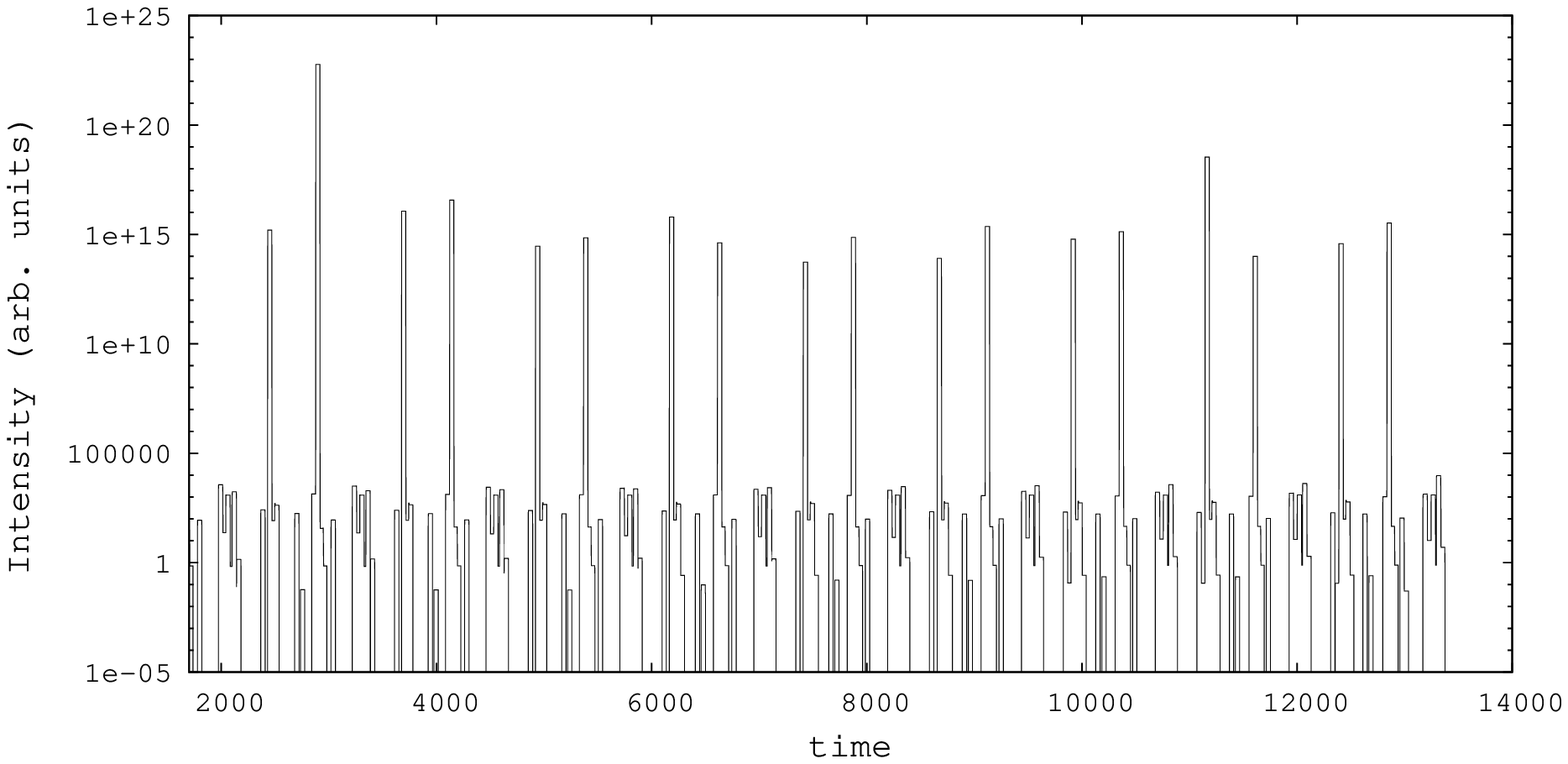}
 \includegraphics[width=3in]{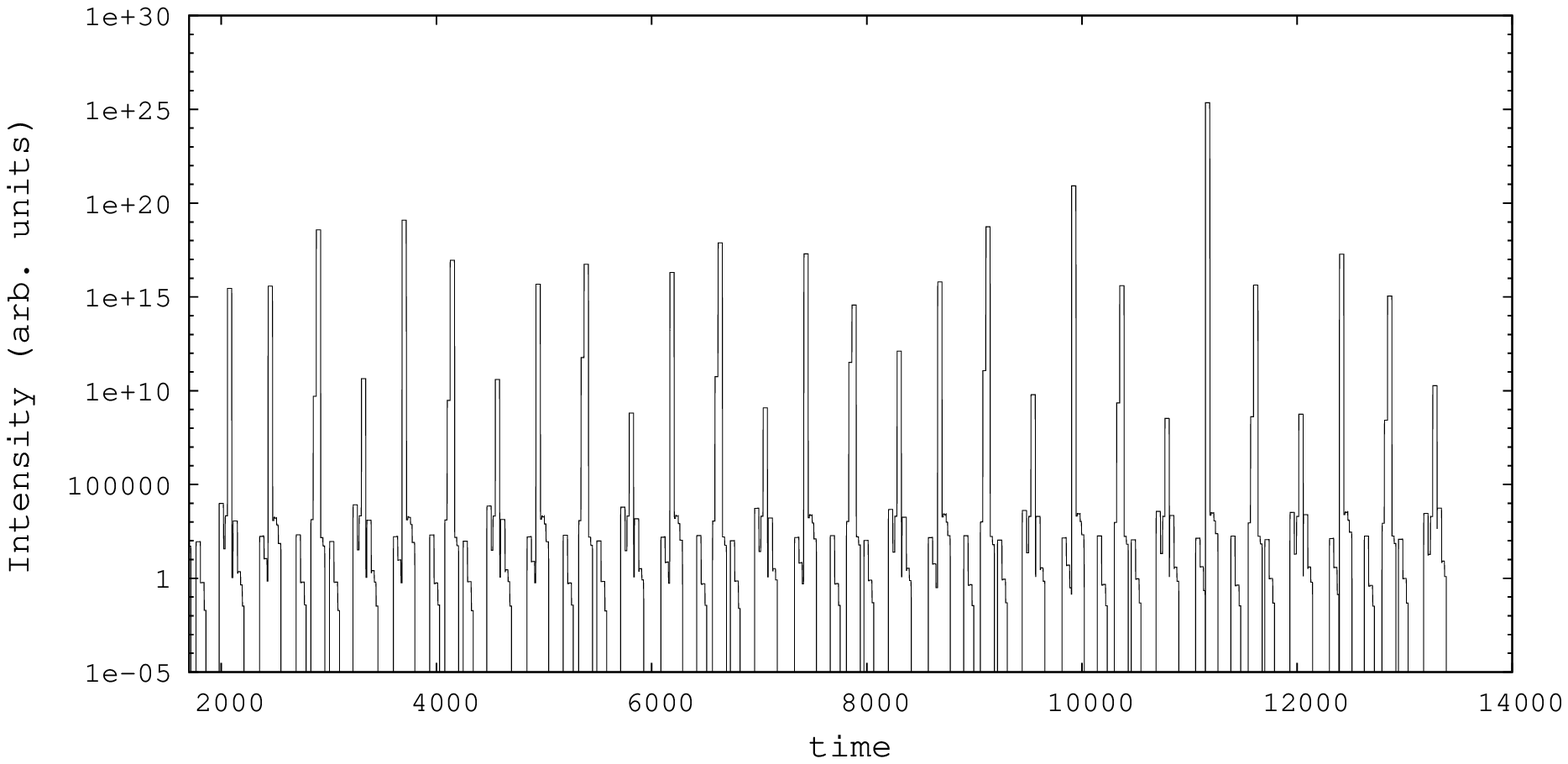}
 \includegraphics[width=3in]{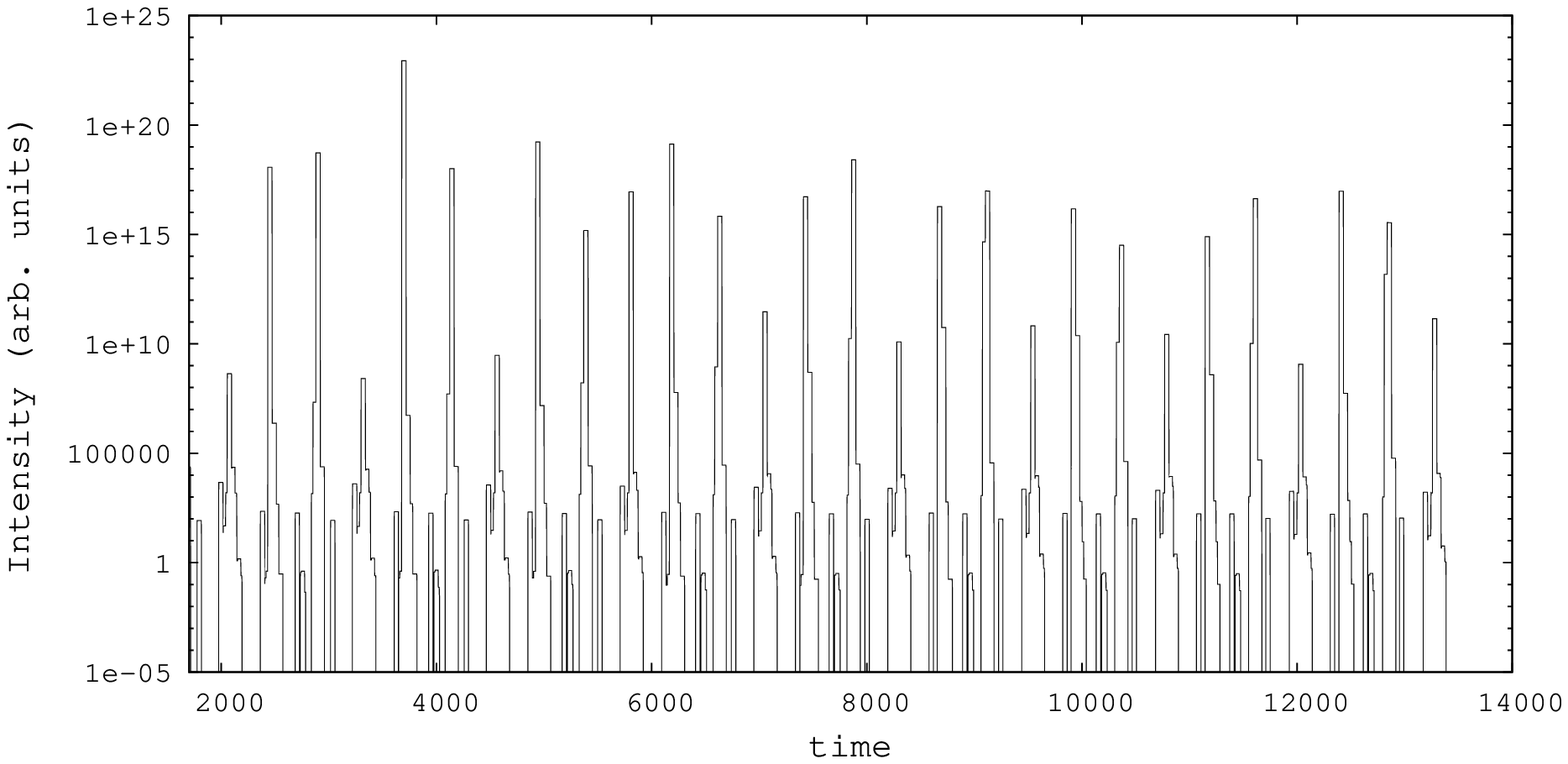}
 \caption{The associated orbits are the same as in the previous figure, but the observing angle is changed to $45^{\circ}$. Light travel time delay makes the observed time between flares uneven when the flares come from the front and back of the black hole. Also, the effects of Doppler boosting are enhanced.} 
 \label{longitude45}
\end{figure}

The corresponding fluxes with a $5^{\circ}$ (top) and $45^{\circ}$ (side) observing angle are plotted in Figs. \ref{longitude5} and \ref{longitude45}. In the first figure, we find the time interval of any double peak is nearly independent of its set-up. However, in the second figure we see that the intervals between peaks are altered. This is because when we are observing from above, the longitude of the ascending node does not affect the propagation time of the signals due to circular symmetry. However, when we are observing from the side, the longitude is an important factor in determining the time delays. The flares behind the black hole need more propagation time than those in front -- the Roemer delay. The order of this time correction is comparable with the time correction due to the precession of the orbit, thus causing the noticeable change in the duration of each double peak. Also, the Shapiro delay times are different, which is caused by the light propagation in curved space-time taking a slightly different path than in flat space-time. Both time delays are the same, however, for signals from the sides due to symmetry since it is non-spinning.

\subsection{Orbital inclination $i$ and spin of the black hole $\tilde{S}$}
The inclination of the stellar orbit is relevant only when the spin of the black hole is large. For a Schwarzschild black hole, the inclination plays no role in determining the quasi-periodicity or the flux of the signals. Also, the effect of spin of the black hole is not prominent unless the orbit is highly inclined.

The spin makes the orbit less periodic via the Lense-Thirring effect, and it no longer is confined to a plane. However, the change to the flare time-series is relatively simple. The result is similar to that from a non-spinning black hole, with the frame dragging inducing an extra modulation of the flare fluxes and timing. This can be seen by comparing the fluxes in Figs. \ref{size} and \ref{spineffect}.

\subsection{Observer's inclination angle $\theta$ and argument of the peribothon $\omega$}
\begin{figure}
 \centering
 \includegraphics[width=3in]{pics/traj/hit_13_5.ps}
 \includegraphics[width=3in]{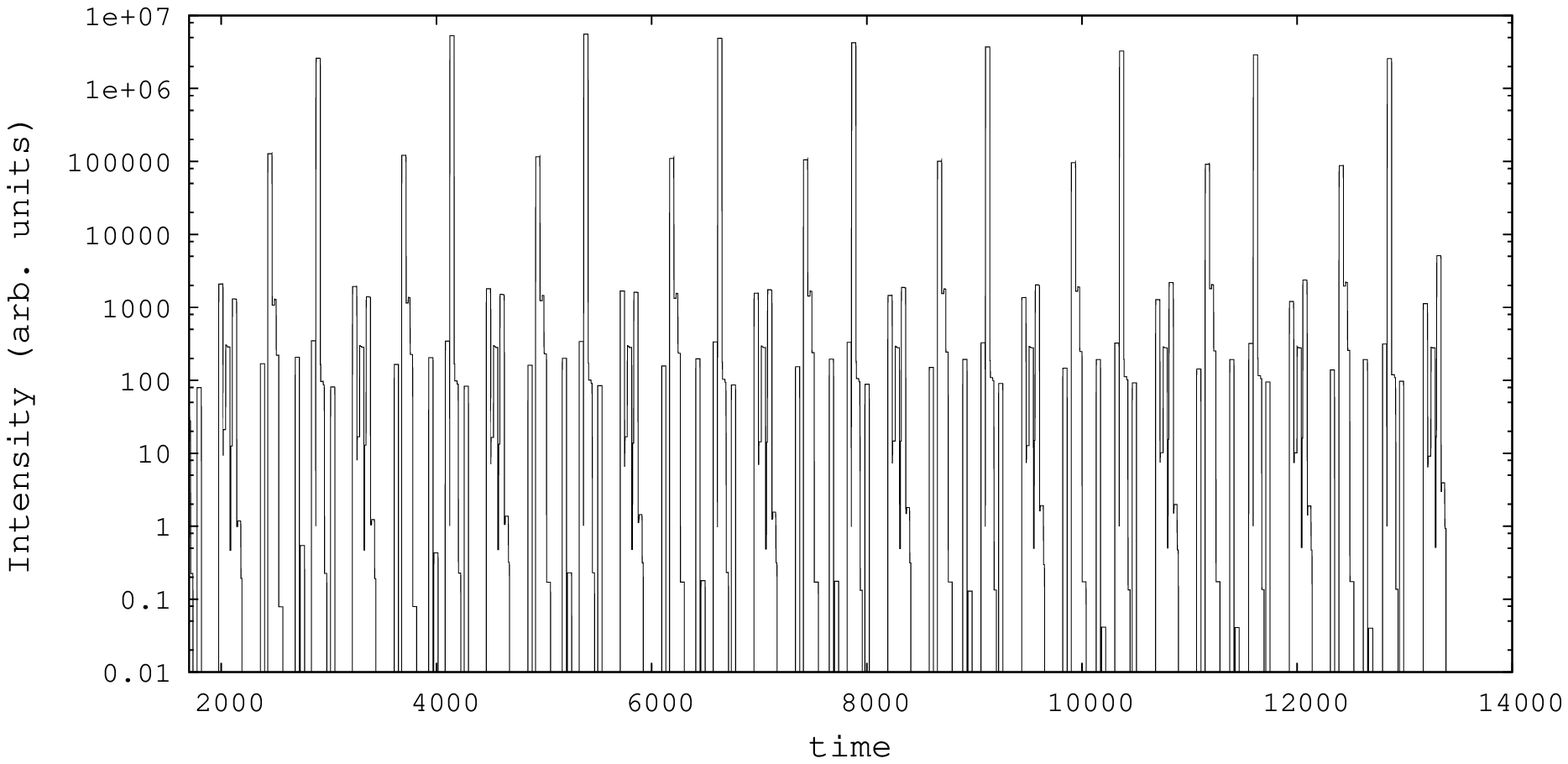}
 \includegraphics[width=3in]{pics/traj/hit_13_45.ps}
 \includegraphics[width=3in]{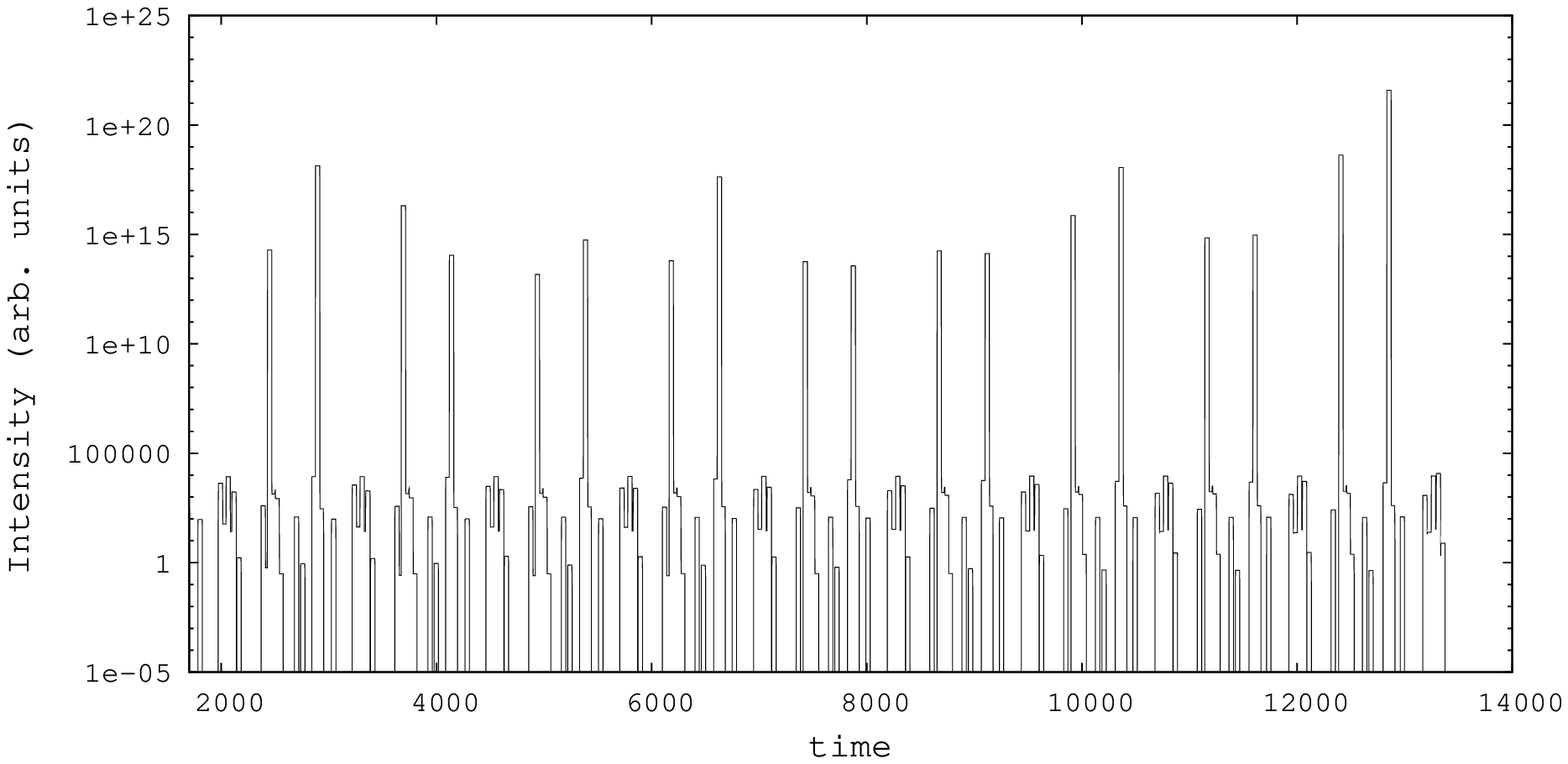}
 \includegraphics[width=3in]{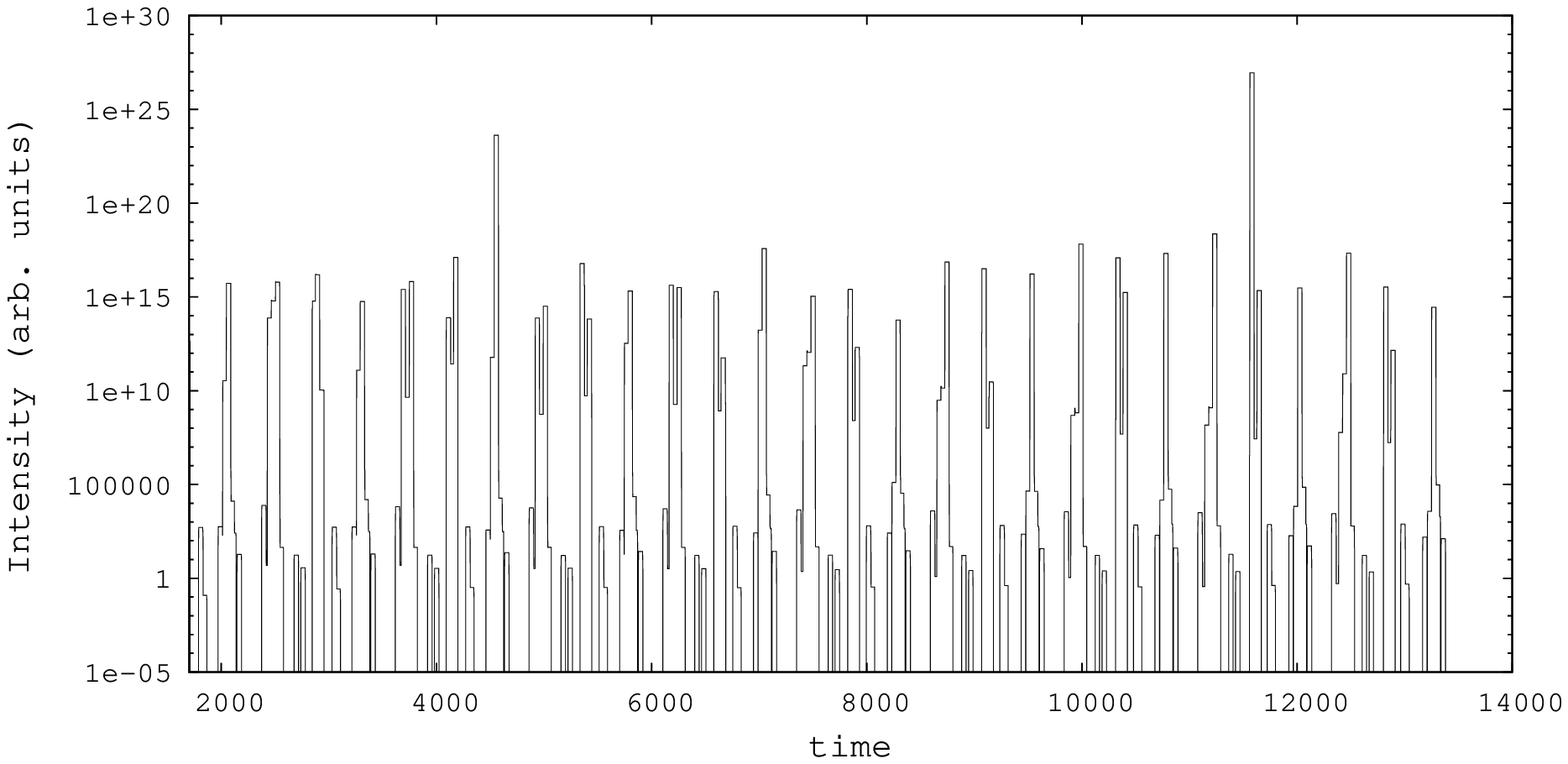}
 \caption{This figure shows how the relative strength of the signals change with the observer's inclination angle. We used the star's orbit in Fig. \ref{a0orbit}b with a Schwarzschild black hole in the center. The five figures correspond to observer's inclination angles of $5^{\circ}$ (from the top), $30^{\circ}$, $45^{\circ}$, $60^{\circ}$, and $85^{\circ}$ (from the side).  As shown in the figures, when observing from the top, the scales of the fluxes are relatively similar, while as the inclination angles of the observer increases, some fluxes' scales start to be increased or decreased. When we are observing from the side, some signals are greatly magnified.} 
 \label{obsinclination}
\end{figure}

\begin{figure}
 \centering
 \includegraphics[width=3in]{pics/traj/hit_14_5.ps}
 \includegraphics[width=3in]{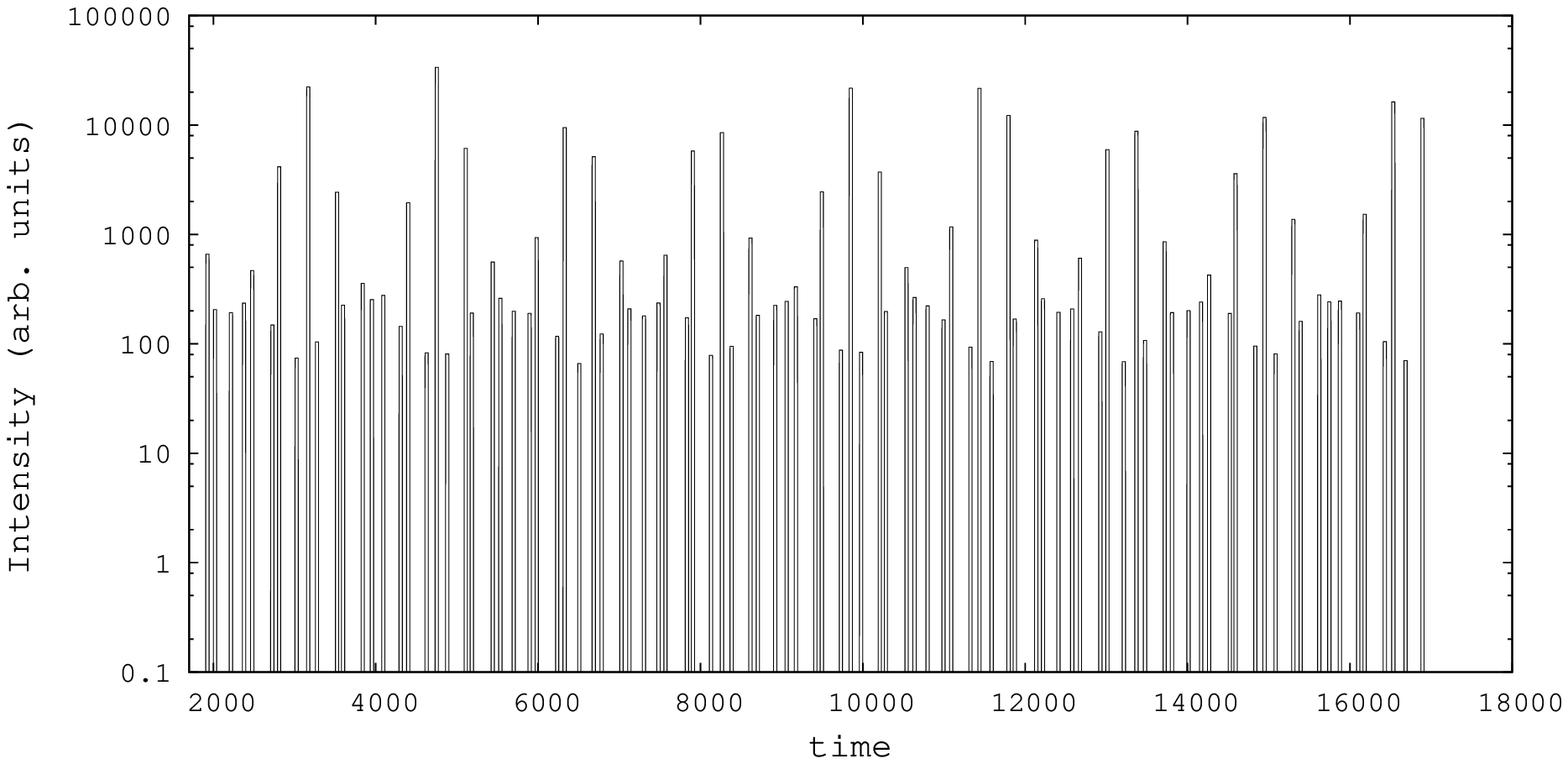}
 \includegraphics[width=3in]{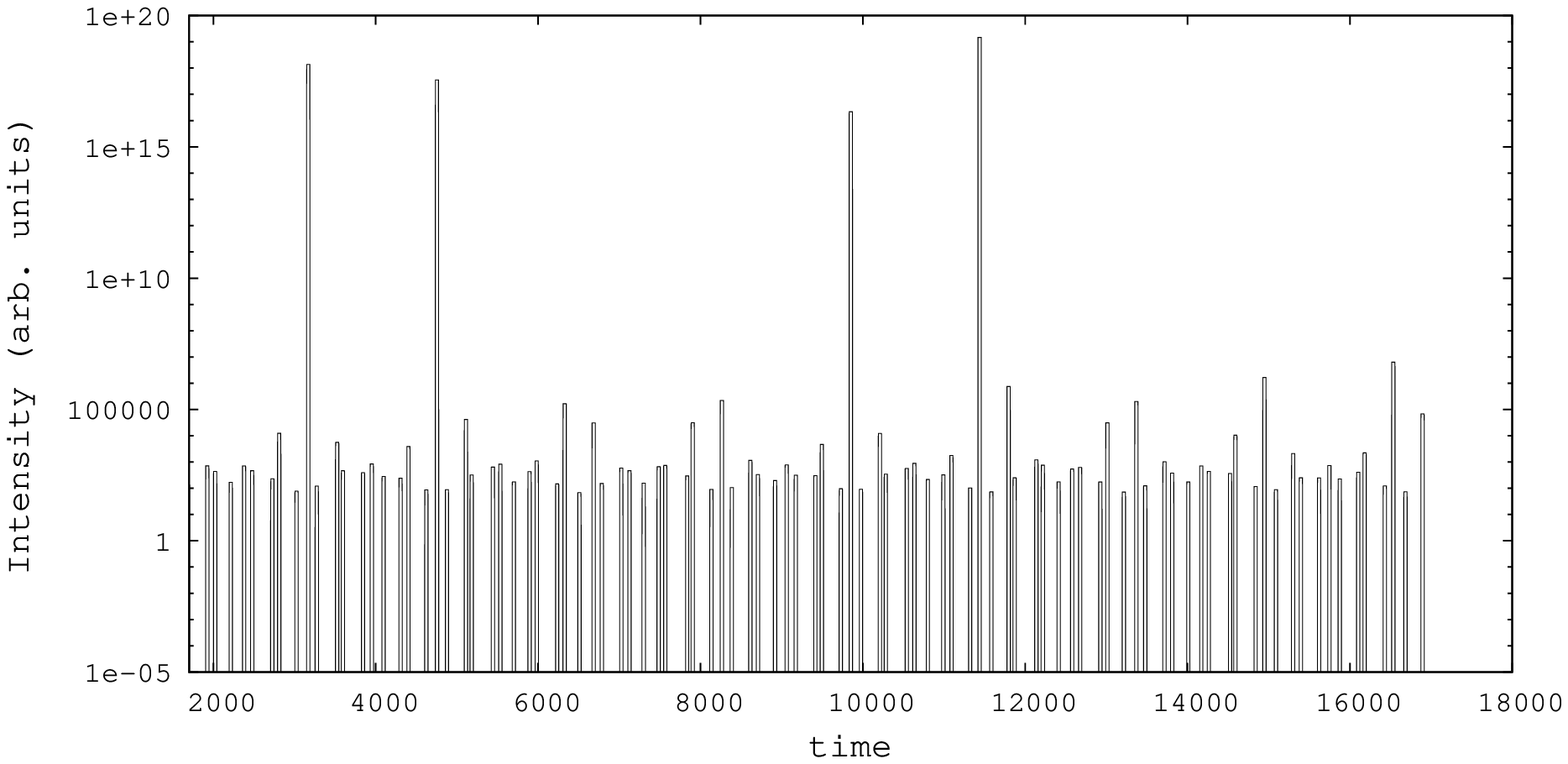}
 \includegraphics[width=3in]{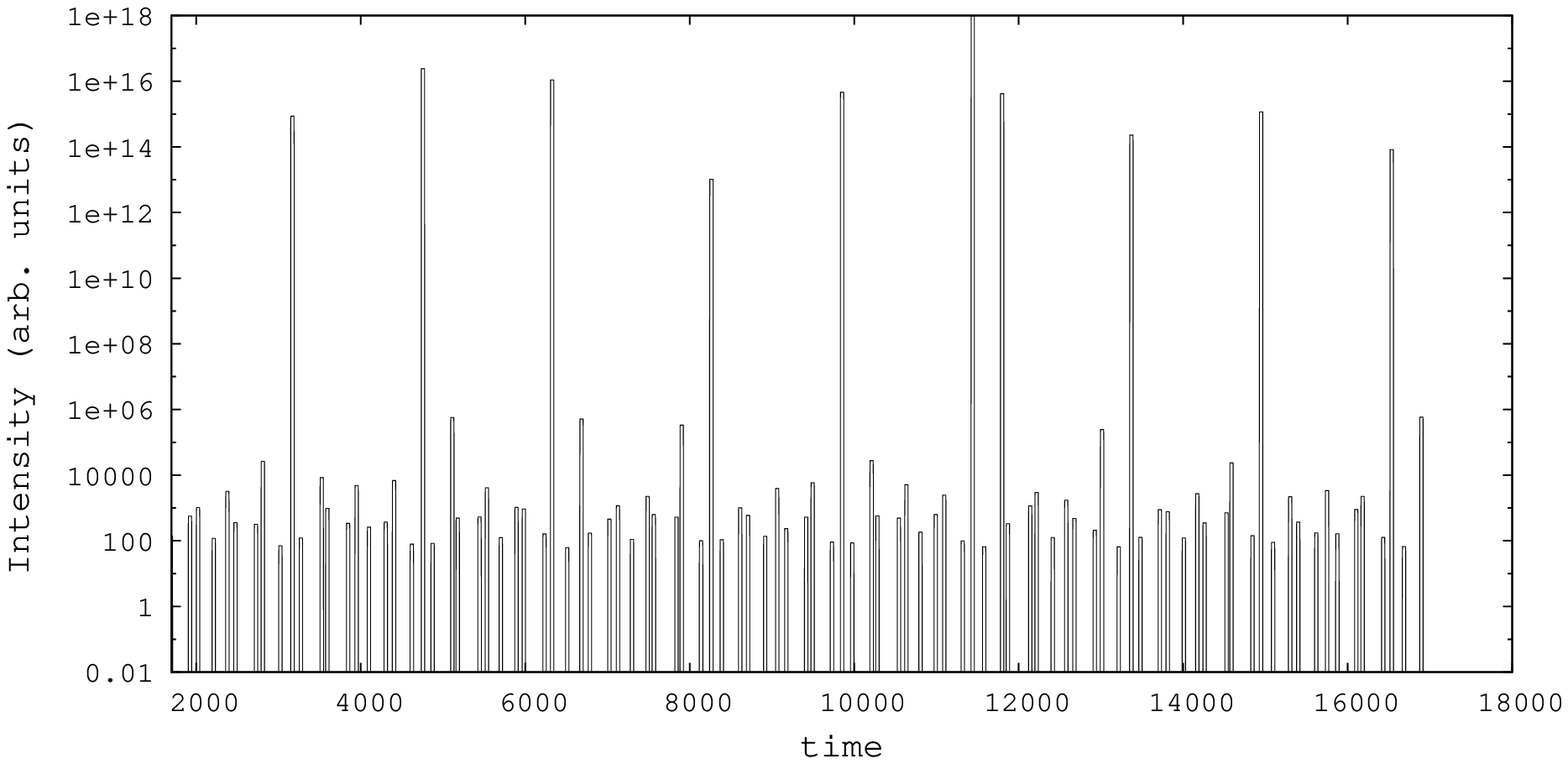}
 \includegraphics[width=3in]{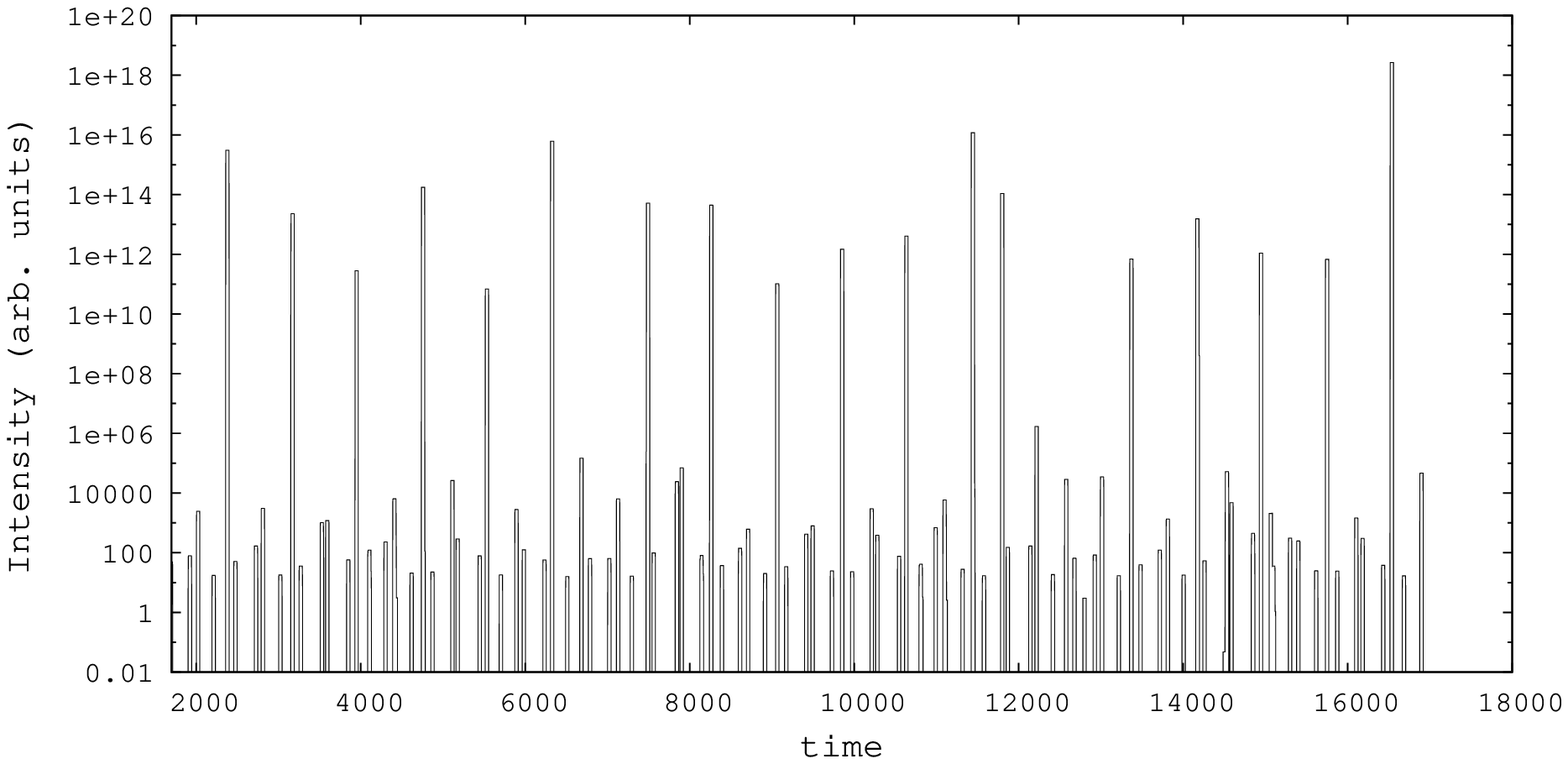}
 \caption{This figure shows the same as Fig. \ref{obsinclination}, but using \ref{a0.998orbit}b and a rapidly spinning black hole.  The higher the inclination, the larger the maximal Doppler boosting effect for the correctly aligned flares.} 
 \label{obsinclinationkerr}
\end{figure}

These parameters only marginally affect the temporal behaviour via differences in the Shapiro and Einstein time delay. However, they can greatly affect the flux intensity. Figs. \ref{obsinclination} and \ref{obsinclinationkerr} show how the strengths of fluxes change as the observer's inclination angle changes. When the inclination is large, some signals are greatly magnified via Doppler boosting, depending on their locations on the accretion disc.

The total energy emitted by the flare depends on the velocity of the star in the gas frame and also upon the relative direction to the line of sight. If one assumes models of accretion disc structure and flare physics, this provides extra information either to determine the parameters of the disc, or to constrain the orbit of the star since Doppler boosting is a sensitive function of angle.

Also note that if two have similar parameters except for the argument of peribothon, flares closer to the peribothon will be more intense intrinsically. This is because the star is moving at a higher speed there. This results in larger variability in the flux intensity profile.

\clearpage

\section{From the signals to the parameters}

The parameters we would like to measure from observations of star-disc collisions are $M, S, a_r, e, \Omega, \omega, i, T_0$, and $\theta$. However, as we discussed in the previous section, under different circumstances, some of these observables are degenerate: they are coupled and cannot be determined individually. Here we study analytically which parameters or combinations of the parameters could be determined from the observations of quasi-periodicity of the signals for the Schwarzschild black hole case.

\subsection{Slowly spinning hole}
In this case we can describe the space-time using the Schwarzschild metric. \citet{Blandford:76}, \citet{Damour:85}, \citet{Damour:86}, \citet{Damour:92}, and \citet{Damour:93} explicitly calculated the post-Newtonian timing formula of binary systems. We employ their formulae since our binary system composed of a Schwarzschild black hole and an orbiting star also behave as a binary system in the moderately strong-field regime of relativistic gravity. They calculated the time delays of the arrival time of signal as:
\begin{equation}
   D\cdot \tau_a = T + \Delta_R(T) + \Delta_E(T) + \Delta_S(T)+  O\left(\left(\frac{v}{c}\right)^4 P\right)
\end{equation}  
up to an arbitrary additive constant, with $T$ as the intrinsic emission time, and each physical quantity is evaluated at $T$.\\
(1) Roemer Time Delay:
\begin{equation}
   \Delta_R(T) =-\hat{k} \cdot \vec{r}_\star,
\end{equation} 
where $\vec{r}_\star$ is the coordinate postion vector from the star to the barycenter of the binary system, which is the SMBH since we assume its mass is much bigger than that of the star, and $\hat{k}$ is the unit vector opposite to the light of sight.\\
(2) Shapiro Time Delay:
\begin{equation}
   \Delta_S(T) = -2M \log{[\vec{k} \cdot (\vec{r}_\star-\vec{r}_{BH}) + |\vec{r}_\star-\vec{r}_{BH}|]}, 
\end{equation} 
where $\vec{r}_{BH}$ is the position vector from the black hole to the barycenter of the binary system, which, in this case, is 0.\\
(3) Einstein Time Delay:
\begin{equation}
    \Delta_E(T) =\gamma \sin{u}.
\end{equation} 
Here we have $\gamma = 2Me/na_r$, with the mean motion $n = \sqrt{M/a_r^3}(1-9M/2a_r)$, and the eccentric anomaly $u$ satisfies $n(t-T_0) = u-e\sin{u}$, with $t$ as the emission time of the signal.

The star hits accretion disc when:
\begin{equation}
    -\hat{k} \cdot \vec{r}_\star = r_\star \cos{\alpha} = \pm r_\star \cos{\theta} \cos{\Omega},
\end{equation}
where $\alpha$ is the angle between the line of sight and $\vec{r}_\star$.  The sign is different for the ascending or descending node.

At the ascending node, $\vec{r}_\star$ makes an angle $\omega$ with the peribothon. Therefore we have:
\begin{equation}
   0 = \omega + (1+\dot{\omega}) A_{e_{\theta}} (u),
\end{equation}
where $e_{\theta}$ is a constant with the value $e(1+2M/a_r)$, and $A_{e_{\theta}}(u)$ is defined as:
\begin{equation}
  A_{e_{\theta}} (u):= 2 \arctan \left[ \left( \frac{1+e_{\theta}}{1-e_{\theta}} \right)^\frac{1}{2} \tan{\frac{u}{2}} \right]
\end{equation}
with the property:
\begin{equation}
  \cos \left(A_{e_{\theta}} (u)\right) = \frac{\cos{u}-e_\theta}{1-e_\theta \cos{u}}.
\end{equation}
Therefore after calculation:
\begin{equation}
  \cos{u} = \frac{\cos\left(\frac{\omega}{1+\dot{\omega}}\right)+e_\theta}{1+e_\theta \frac{\omega}{1+\dot{\omega}}},
\end{equation}
and
\begin{equation}
   r = \frac {a_r(1-e_\theta^2)}{1+e_\theta \cos{\left(\frac{\omega}{1+\dot{\omega}}\right)}}.
\end{equation}
Similarly, at the descending node, 
\begin{equation}
   r = \frac {a_r(1-e_\theta^2)}{1-e_\theta \cos{\left(\frac{\omega}{1+\dot{\omega}}\right)}}.
\end{equation}
Thus, we obtain:
\begin{equation}
    - \vec{k} \cdot \vec{r} = \mp \frac{a_r(1-e_\theta^2)}{1\pm e_r\cos{\left(\frac{\omega}{1+\dot{\omega}}\right)}} \cos{\theta} \cos{\Omega},
\end{equation}
where $-$ stands for the ascending, and $+$ stands for the descending node.
Also, using
\begin{equation}
  \sin \left(A_{e_{\theta}}(u)\right) = \frac{\sin{u}\sqrt{1-e_\theta^2}}{1-e_\theta \cos{u}},
\end{equation}
we obtain:
\begin{equation}
  \sin{u} = \mp \frac{\sqrt{1-e_\theta^2}}{1\pm e_\theta \cos\left(\frac{\omega}{1+\dot{\omega}}\right)}\sin{\left(\frac{\omega}{1+\dot{\omega}}\right)}. \\
\end{equation}
Therefore, the time delays are
\begin{equation}
   \Delta_R(T) = \mp \frac{a_r (1-e_\theta^2)}{1\pm e_\theta \cos{\left(\frac{\omega}{1+\dot{\omega}}\right)}} \cos{\theta} \cos{\Omega},
\end{equation} 

\begin{equation}
   \Delta_S(T) = - 2M \log \left[\frac{a_r (1-e_\theta^2)}{1\pm e_\theta\cos\left(\frac{\omega}{1+\dot{\omega}}\right)} \left(1\pm \cos{\theta}\cos{\Omega}\right)\right], 
\end{equation} 
and 
\begin{equation}
    \Delta_E(T) =\mp \frac{2Me}{na_r} \frac{\sqrt{1-e_\theta^2}}{1\pm e_\theta \cos\left(\frac{\omega}{1+\dot{\omega}}\right)}\sin{\left(\frac{\omega}{1+\dot{\omega}}\right)}.\\
\end{equation}

Define $\alpha$ as the angle between the line connecting the SMBH to the observer and the line connecting the SMBH to the ascending node, then $\cos{\theta} \cos{\Omega} = \cos{\alpha}$. We cannot decouple $\theta$ and $\Omega$, since the system is rotationally invariant about the line at which the stellar orbital plane intersects the accretion disc. In addition, define $\bar{\omega}=\omega/(1+\dot{\omega})$. $\bar{\omega}$ equals the initial argument of peribothon $\omega$ if there is no precession. Then the three time delays become:

\begin{eqnarray}
   \Delta_R(T) &=& \mp a_r  \frac {1-e^2(1+q)^2}{1\pm e(1+q) \cos{\bar{\omega})}} \cos{\alpha}\ ,\nonumber\\
   \Delta_S(T) &=& - a_r q \log \left[\frac {T_\star (1-e^2(1+q)^2)}{1\pm e(1+q)\cos{\bar{\omega}}}(1\pm \cos{\alpha})\right]\ ,\nonumber \\
   \Delta_E(T) &=&\mp a_r \frac{e\sqrt{2q}}{ \left(1-\frac{9q}{4}\right)} \frac{\sqrt{1-e^2(1+q)^2}}{1\pm e(1+q) \cos{\bar{\omega}}}\sin{\bar{\omega}},
\label{timed_eqns}
\end{eqnarray}
where  $q\equiv 2M/a_r$ is dimensionless quantity, which compares the scale of the black hole and the scale of the stellar orbit. In addition, the precession rate of peribothon and the orbital frequency become

\begin{eqnarray}
  \dot{\omega} &=& \frac{3q\sqrt{2q}}{4a_r (1-e^2)}\ ,\nonumber \\
  \omega_k &=& \frac{\pm\sqrt{\frac{q}{2}}}{a_r }\ .
\label{prec_eqns}
\end{eqnarray}
These define the observables obtainable from pure timing measurements.

Theoretically, one would only require six flares to solve for the reasonable parameters of the system: $T_\star, q, e, \cos \alpha$, and $\bar{\omega}$.  There are five unknowns, and five equations via (\ref{timed_eqns}) and (\ref{prec_eqns}) for them, together with an extra equation/unknown for the time initialization $T_0$.  These theoretically can be measured and simultaneously solved to determine the system.

Unfortunately, the above equations cannot be directly used to calculate the parameters of the system from observations. They are only approximations, and do not include the effects of the black hole spin on the light delay time. In addition, they assume that the orbit is an ellipse which is incorrect close to the black hole.  Note that the orbital trajectory can be integrated analytically in terms of elliptic functions \citep{Fanton:97}. However, these expressions cannot be easily inverted. Thus, a solution will require some form of numerical fitting of the observations and more than the minimal number of flares as data points.

The best approval appears to be hierarchical fitting. Some parameters are much easier to determine than others. By looking at the modulation of the flare signals, it should be possible to obtain the period and precession rates. With these, it seems to be possible to reduce the size of the multidimensional space enough that a direct search of possibilities for the remaining parameters is possible.

\subsection{Rapidly spining holes}

In this case, we must use the full Kerr metric, and only limited progress can be made analytically. An analysis of observational data would undoubtably involve numerical integration of the orbit and ray equations.

When we add the spin to the group of the parameters, one more equation needs to be added to solve the problem in principle, which is the Lense-Thirring precession rate. The precession rate of orbital frequency is:
\begin{equation}
  \dot{\Omega} = \frac{\tilde{S} q^2}{2 T_\star (1-e^2)^\frac{3}{2}}\ ,
\label{prec_LT}
\end{equation}
The other five equations need to have corrections due to the spin too, but we can express two of them analytically:
\begin{eqnarray}
  \dot{\omega} &=& \frac{1}{T_\star} \left(\frac{3q\sqrt{2q}}{4 (1-e^2)} - \frac{q^2 \tilde{S}}{(1-e^2)^\frac{3}{2}} \right)\ ,\nonumber \\
  \omega_k &=& \frac{\pm1}{T_\star  \left(\sqrt{\frac{2}{q}} \pm \frac{q}{2} \tilde{S} \right)}\ ,
\label{prec_eqns_kerr}
\end{eqnarray}
while the other three time-delay formulae are extremely hard to calculate analytically. Fortunately, we have another way to estimate the spin parameter besides using numerical methods.

As we have shown in Fig. \ref{spineffect}, the spin of the black hole would transfer power to higher harmonics in the time series, which increases the difficulty of studying the spin of the hole through the timing analysis. However, due to the precession of the longitude of the ascending node and descending node, flares triggered at different locations of the disc would have different Doppler shifts in their intensity. Since the peribothon advance affects the intensity change marginally, we have a reason to believe that the flare intensity modulation is basically caused by the Lense-Thirring effect and the period of this modulation is related to the spin of the black hole.

From \citet{Ciu:95}, we have: 
\begin{equation}
  \tilde{S} = \frac{a_r^3 (1-e^2)^\frac{3}{2}}{2M^2} \dot{\Omega}.
\end{equation}
If we define the Lense-Thirring precession period to be $P_\Omega = 2 \pi / \dot{\Omega}$,
\begin{equation}
  \tilde{S} = \frac{\pi a_r^3 (1-e^2)^\frac{3}{2}}{M^2 P_\Omega} .
\end{equation}

Tests using the modulation period of the flux in Fig. \ref{spineffect} and the size and eccentricity of the stellar orbit from the simulated orbits in Fig. \ref{a0.998orbit} have been performed in which the spins are all calculated to be correct in order of magnitude. Therefore, the period of the modulation of the flare fluxes, together with the size and shape of the stellar orbit, and the mass of the black hole inferred by the orbital period, can provide a probe into the spin of the hole, though more studies need to be done on how the spin can be calculated more accurately directly.

\section{Observations}
\subsection{X-ray flares from tidal disruption of stars near slowly accreting SMBHs}

Stars on plunging orbits in a galactic core may approach close enough to the SMBH to lose mass \citep{Rees:88, Evans}. A fraction of the outflowing gas may be captured and accreted by the black hole, producing outbursts of radiation.

\citet{Komossa:99} reported non-recurrent X-ray flares from NGC 5905 and IC 3599. The former is an inactive galaxy, and a tidal disruption event with peak luminosity exceeding $L_x = 10^{42}$ erg s$^{-1}$ fits the data best. Using the Eddington luminosity, \citet{Komossa:99} calculated that the lower bound of the SMBH mass would be $10^5 M_\odot$.  This outburst lasted for several months, consistent with the theoretical calculation of \citet{Rees:90}.

Moreover, \citet{Komossa:01} reported a few more X-ray flares from inactive galaxies, all of which ould be interpreted as tidal disruption events.  These soft X-ray flares have large luminosities, high degrees of variability, and the absence of optical signs of Seyfert activity. \citet{Komossa:08} further discussed this scenario in the environment of recoiling SMBHs.

\subsection{X-ray Periodicity from the AGN in galaxy RE J1034+396}

\citet{Gierlinski:08} reported the observed periodic XMM-Newton X-ray signals from a nearby active Seyfert I galaxy RE J1034+396.  As shown in Fig. 1 in \citet{Gierlinski:08}, the signals have a periodicity around 1 hour (more precisely, at $3730\pm130$ s). The X-ray flux has a modulation of around $10\%$ and is identified with the nucleus on bolometric grounds.

The data were divided into two segments. For the first seven periods, there are phase shifts and extra maxima. Then after $t_0 = 25$ ks,  in the second segment, there are about 16 coherent cycles. \citet{Gierlinski:08} folded the second segment and showed that the X-ray modulation could be well fitted by a cosine function. 

We consider two scenarios to explain the observed X-ray periodicity and sinusoidal light curve. The first is that the star orbits equatorially inside a thick disc, and emits X-rays due to its velocity relative to the disc. The relative velocity might result from the fact that the disc will be pressure-supported and sub-Keplerian, the gas within the disc can have an inward velocity, or the star orbits with a nontrivial eccentricity. The rapid inflow would preclude the creation of a gap in the disc. Another scenario is that instead of having photons produced directly by the star-disc interaction, we can have a star slowly overflowing its Roche lobe as it orbits. This would be a source of accreting gas and the modulated emission may be produced at a variable hotspot where the accretion stream from the Lagrange point hits the disc. If this is the case, predictions of the the evolution of the stellar orbit can be made. Further discussion of this model and RE J1034+396 will be described elsewhere.

\subsection{Periodic optical outbursts of OJ287}
The BL Lac object OJ 287's optical outbursts exhibit periodicity of roughly 12 years (\citet{Nilsson:06}, and references therein), as shown in Fig. 1 in \citet{Nilsson:06}. The last two outbursts also show a double-peak feature of roughly a year separation.

Many models have been suggested, and one of them proposed by \citet{Lehto:96} suggests that these periodic flares can be fitted by the orbit of a secondary $10^8 M_\odot$ SMBH around a primary $1.7 \times 10^{10}M_\odot$ SMBH. The orbital period is 12.07 years, and two flares are produced when the secondary SMBH crosses the accretion disc of the primary one due to thermal bremsstrahlung. As shown in Fig. 2 in \citet{Nilsson:06}, the time separation between flare 1 and 2 is roughly a year. This model is equivalent to ours with a $1.7 \times 10^{10} M_\odot$ Schwarzschild central black hole. Since the secondary black hole is relatively far from the center, the orbit is nearly periodic. Here the orbit has a precession of about $33^\circ$ per orbit. More observations will be necessary to see if this model is highly predictive.

\subsection{Infrared flares from the Galactic center}
Our Galactic center hosts a supermassive black hole of mass $4.1\sim4.5 \times 10^6$ $M_\odot$ (\citet{Genzel:03, Ghez:08, Gillessen}; and references therein), which gives a gravitational radius $R_g=M \sim 18$ light seconds. The innermost stable orbit, which ranges from 1.24 to 6 $R_g$ depending on the unkown spin of the black hole, will be about $0.1\sim1$ light minute. The radio size of the black hole, as detected by VLBI \citep{Bower, Shen}, is less than 10 light minutes.

Two groups mapped the central stars in the Galactic center, as in Fig. 2 in \citet{Genzel:07}. The closest observed stellar orbit peribothon is about 17 light hours in radius \citep{Schodel:02, Schodel:03, Ghez:03, Ghez:05, Eisenhauer:05}. These observable stars are bright OB stars. For a given supermassive black hole, the tidal radius $R_T$ of an orbiting star is proportional to $R_\star / {{M_\star}^{1/3}}$ \citep{Rees:88}. In the lower mass range of main sequence stars, we have $R_\star \propto M_\star ^{0.8}$ \citep{Kippenhahn}. Therefore, for main sequence stars $R_T \propto M_\star^{0.47}$. This means the less massive a star is, the denser it is, and the closer it can get to the central black hole. Therefore, a sun-like star can get as close as around $5$ light minutes to the center of the black hole in the Galactic center; a dwarf star can get to around $10 R_g$; white dwarfs and neutron stars have higher densities than main sequence stars, and will cross the horizon intact.

\citet{Genzel:03} detected quasi-periodic infrared flares using ESO-VLT, as shown in their Fig.2. The signals have a 17 min period between consecutive peaks and a 1 day period between groups of peaks. If we try to explain these flares with our star-disc collision model, we need to invoke a highly eccentric stellar orbit around a Kerr SMBH with an orbital period $\sim$ 1 day. This period would allow almost all stars to escape from tidal disruption. However, we do not have any observation of stars that close to the hole, and there are 3 -- 5 consecutive peaks in each group, which we cannot explain with our model. Suppose a hot spot is produced in each collision and it orbits around the hole for several cycles. The consecutive peaks should then follow a power-law curve instead of the rather symmetric distribution in Fig. 2e in Genzel's paper. Therefore, a star-disc collision model seems an unpromising explanation of this quasi-periodicity.

\section{Discussion}

In this paper, we have explored observable effects that can arise when a normal star follows a bound orbit close to the event horizon of a massive black hole in a galactic nucleus. We have paid special attention to the quasi-periodic signal that may be detected in X-ray observations if the star creates an X-ray pulse every time it crosses the disc plane. We have also considered, briefly, alternative forms of signal modulation, especially associated with more subtle star-disc interactions involving hot spots and tidal shocks \citep{Spruit:87}. If the orbit is relativistic, the Doppler effects will almost guarantee a measurably modulated signal.

Although the detection of effects like this would be of considerable interest in its own right, it would open up two more far-reaching lines of inquiry. The first of those concerns the testing of general relativity in the strong field regime. To date, all quantitative tests of relativity have been executed in the weak field regime. (We exclude interpretations of observations such as asymmetrically broad iron fluorescence lines, e.g. \citet{Miller:07}, and QPOs in stellar black hole discs, e.g. \citet{Bob:99, Bob:01}.) We have described the types of time series that could be measured and shown how these can be interpreted initially as measurements of orbital and black hole parameters and then, given sufficient high-quality data, as prescriptive tests of the general relativistic calculation of orbits in a Kerr space-time. Such a test would represent a major advance, though it would not be as powerful as the measurement of a gravitational radiative waveform from a major black hole merger.

The second line of inquiry is that observations would allow us to quantify the rate of extreme mass-ratio inspirals (EMRI) in galactic nuclei. (The discovery of ~100 short-lived OB stars in the Galactic center underscores how little we understood about the co-existence of stars and massive black holes in galactic nuclei.) This would, in turn, relate to understanding the role of stars in the behavior of massive accretion discs. It would also relate most directly to the planning for observations such as LISA that are designed to detect gravitational signals from stellar inspirals.

Of course, most exciting of all would be to understand a stellar orbit well enough that the stars ``quietus" -- either at the ISCO or through rapid tidal destruction -- could be predicted and observed. Given our current ignorance, we cannot rule out such an opportunity arising within our lifetime.

\section*{Acknowledgments}

This work is supported in part by the U.S. Department of Energy contract to SLAC No. DE-AC3-76SF00515. We would like to thank R. Wagoner and S. Healey for helpful discussions. 


\bsp

\end{document}